\documentclass[aps,prc,twocolumn,fleqn]{revtex4-1}

\usepackage{amsmath, amssymb}
\usepackage[pdftex]{graphicx}
\usepackage{color}

\DeclareMathOperator{\tr}{Tr}

\begin{document}

\title{Early Time Dynamics of Gluon Fields in High Energy Nuclear Collisions}

\author{G. Chen}
\affiliation{Department of Physics and Astronomy, Iowa State University, 
Ames IA 50011, USA}
\affiliation{Cyclotron Institute and Department of Physics and Astronomy,
Texas A\&M University, College Station TX 77843, USA}
\author{R. J. Fries}
\affiliation{Cyclotron Institute and Department of Physics and Astronomy,
Texas A\&M University, College Station TX 77843, USA}
\author{J. I. Kapusta}
\affiliation{School of Physics and Astronomy, University of Minnesota,
Minneapolis MN 55455, USA}
\author{Y. Li}
\affiliation{Department of Mathematics and Statistics, University of Minnesota
  -- Duluth,
Duluth MN 55812, USA}

\begin{abstract}
Nuclei colliding at very high energy create a strong, quasi-classical gluon field during the initial phase of their interaction. We present an analytic calculation of the initial space-time evolution of this field in the limit of very high energies using a formal recursive solution of the Yang-Mills equations.  We provide analytic expressions for the initial chromo-electric and chromo-magnetic fields and for their energy-momentum tensor. In particular, we discuss event-averaged results for energy density and energy flow as well as for longitudinal and transverse pressure of this system. For example, we find that the ratio of longitudinal to transverse pressure very
early in the system behaves as $p_L/p_T = -[1-\frac{3}{2a}(Q\tau)^2]/[1-\frac{1}{a}(Q\tau)^2]+\mathcal{O}(Q\tau)^4$ where $\tau$ is the longitudinal proper time, $Q$ is related to the saturation scales $Q_s$ of the two nuclei, and $a = \ln (Q^2/\hat{m}^2)$ with $\hat m$ a scale to be defined later.  Our results are generally applicable if $\tau \lesssim 1/Q$.  As already discussed in a previous paper, the transverse energy flow $S^i$ of the gluon field exhibits hydrodynamic-like contributions that follow transverse gradients of the energy density $\nabla^i \varepsilon$. In addition, a rapidity-odd energy flow also emerges from the non-abelian analog of Gauss' Law and generates non-vanishing angular momentum of the field. We will discuss the space-time picture that emerges from our analysis and its implications for observables in heavy ion collisions. 
\end{abstract}

\maketitle

\section{Introduction}

Collisions of nuclei at high energy at the Relativistic Heavy Ion Collider (RHIC) and the Large Hadron Collider (LHC) have established the existence of a deconfined phase of partons at high energy densities $\varepsilon \gtrsim 1$ GeV/fm$^3$ \cite{Adcox:2004mh, Adams:2005dq,Muller:2012zq}. The future goal of these programs is to  make precision measurements of properties of quark gluon plasma (QGP) and to study further details of the phase diagram of quantum chromodynamics (QCD). This ambitious task requires a detailed understanding of the bulk dynamics in nuclear collisions. The most promising candidate theory for understanding the initial phase of these collisions is color glass condensate (CGC) \cite{McLerran:1993ka,McLerran:1993ni,Iancu:2003xm,Gelis:2010nm} in which the initial interaction of nuclei, modeled as a collection of $SU(3)$ color charges before the collision, leads to a quasi-classical gluon field after the collision. This field eventually decays into a thermalized QGP.

Once the system is close to local kinetic equilibrium, dissipative relativistic fluid dynamics has become the tool of choice to compute the expansion and cooling of the QGP fireball. Comparisons of hydrodynamic simulations with experimental data have been increasingly successful in pinning down the shear viscosity and the equation of state of high temperature nuclear matter \cite{Kolb:2003dz,Romatschke:2007mq,Teaney:2009qa,Song:2010mg,Gale:2013da}.  The equilibration time $\tau_{\mathrm{th}}$, when hydrodynamic concepts can be applied, as well as the initial values for energy density, energy flow, and all other currents at $\tau=\tau_{\mathrm{th}}$, are often treated as parameters in the fluid dynamic simulation. Model calculations of the initial state, such as the Glauber model \cite{Miller:2007ri}, often constrain only a small subset of initial parameters.  In particular, initial transverse flow is still often poorly constrained in many calculations or even neglected despite very good arguments to the contrary \cite{Kolb:2002ve,Vredevoogd:2008id}.  If color glass condensate is found to be the applicable description of the initial interaction of nuclei at collider energies we will, in principle, be able to calculate the initial conditions at the time $\tau_{\mathrm{th}}$. Recent progress seems to indicate that this is the correct path \cite{Schenke:2012wb,Gale:2012rq}.

Here we have a modest goal. We would like to present analytic results that bridge the gap between known results for the classical gluon field in single nuclei before the collision \cite{JMKMW:96} and the glasma fields at a time $\tau_0 \sim 1/Q$ after the collision. The $\tau_0$ represents the limit of convergence of the small-time expansion we employ. However, in terms of physics it also represents the point at which the longitudinal pressure $p_L$, initially large and negative, approaches zero or even becomes positive, a necessary (but not sufficient) condition for pressure isotropization. Our results then provide solid and urgently needed input to constrain the energy-momentum tensor at a later time $\tau_\mathrm{th} > \tau_0$ which can feed into fluid dynamic simulations. It might be used in an ad-hoc thermalization approximation, as in \cite{Fries:2005yc,Gale:2012rq}, or it might serve as the starting point of further studies of thermalization itself \cite{Gelis:2013rba,Berges:2013eia}. The phenomenology we find is surprisingly rich. For example, the system has non-zero angular momentum and exhibits
directed flow. It resembles aspects of phenomenological models based on QCD strings or string ropes suggested previously \cite{Magas:2000jx,Magas:2002ge}; 
however, our derivation here is based strictly on classical QCD.

We should note that while $\tau_0$ is rather early in the collision, it is within this initial time period that important global properties are set.  These include how much energy, momentum, and angular momentum are transfered from the initial system of colliding nuclei and deposited in the relevant part of the fireball around midrapidity.  While we will focus on analytic results for event-averaged quantities, it is in principle straight-forward to construct a semi-analytic event generator
based on our results.

Color glass condensate has been developed from the idea that nuclear wave functions in the asymptotic limit of very high energies should exhibit novel properties of QCD \cite{McLerran:1993ka,McLerran:1993ni,KoMLeWei:95,Kovner:1995ja,JMKMW:96,Kovchegov:1996ty,Kovchegov:1998bi,Iancu:2003xm,Gelis:2010nm}.  This state is characterized by a slowing growth of the gluon distribution with increasing energy (or decreasing Bjorken-$x$). The gluon area density in a hadron or nucleus saturates and thus defines a saturation scale $Q_s$. We will denote the proper saturation scale in a nucleus by $Q_s$ and will assume it is related to the scale $Q$ used earlier by a numerical factor. We will discuss the ultraviolet scale $Q$ in more detail later.  At high energies $Q_s$ becomes large, $Q_s\gg \Lambda_{\rm QCD}$, and the strong coupling $\alpha_s$ becomes small.  The $Q_s$ is assumed to be on the order of a few GeV in heavy ion collider experiments.  In addition, gluon occupation numbers are large and a quasi-classical description of the gluon field becomes applicable.  If two nuclei collide at high energy, the interaction of the two color glass condensate states create what is sometimes referred to as glasma \cite{KoMLeWei:95,Kovner:1995ja,Fries:2005yc,Fries:2006pv,Lappi:2006fp}.  Here we are interested in the early time evolution of glasma. We will use the classical approximation, known as the McLerran-Venugopalan (MV) model \cite{McLerran:1993ka,McLerran:1993ni,KoMLeWei:95,Kovner:1995ja}.  We will, however, need to generalize the original form of the MV model in this work to allow for a rigorous description of transverse dynamics.  Quantum corrections have been studied and seem to indicate that the classical description is adequate to describe the evolution of the system up to times of order $1/Q_s$ \cite{Gelis:2013rba}. Initial small fluctuations can grow exponentially at times beyond $1/Q_s$ and lead to instabilities. They are probably an important step on the path to thermalization.  Recently, important progress has been made on this phase in the evolution of gluon fields \cite{Gelis:2013rba,Berges:2013eia}. 

The time $\tau_0$ has multiple important implications in our work.  It signals the breakdown of the classical approximation as well as the limit (on purely mathematical grounds) of our specific solution to the Yang-Mills equations. However, it also heralds decoherence of the classical fields \cite{Fries:2008vp} at which the net transfer of energy and angular momentum from the receding nuclei onto the fireball presumably stops, and it is responsible for most of the reduction of the pressure asymmetry (neglecting transverse gradients)
\begin{equation}
  \label{eq:isotrop}
  \frac{p_L-p_T}{(p_L+2p_T)/3} = -6 \left[\frac{1-\frac{5}{4a} (Q\tau)^2}{1-\frac{1}{2a} (Q\tau)^2}\right]
+ \mathcal{O}(Q\tau)^4 \, ,
\end{equation}
where $a = \ln (Q^2/\hat{m}^2)$.  This will be discussed in Sec. \ref{sec:6}.

Our paper is organized as follows. In Sec. \ref{sec:2} we review the MV model for single nuclei and colliding nuclei on the light cone. We discuss a recursive solution of the equations of motion of the gluon field. We also compare the emerging space-time picture to existing phenomenological approaches.  In Sec. III we calculate the energy-momentum tensor of the early gluon field as a function of the initial color electric and magnetic fields up to fourth order in proper time $\tau$. In Sec. IV we generalize the assumptions used to calculate expectation values of observables in the MV model and redo the classical calculation of the gluon distribution function of a
nucleus. We then proceed to calculate the expectation values, or event averages, of gluon field correlation functions of higher twist which will be needed later on.
In Sec. V we compute the expectation value of the glasma energy-momentum tensor up to fourth order in $\tau$, although at third and fourth
order in $\tau$ only leading contributions in $Q$ are computed.  In Sec. VI we explore the phenomenological consequences including pressure anisotropies and flow. Section VII summarizes our results.

\section{The Gluon Field of Two Color Charges on the Light Cone}
\label{sec:2}

In this section we discuss analytic solutions for the Yang-Mills equations of two nuclei colliding on the light cone with non-abelian $SU(N_c)$ charges kept fixed.  The setup is reminiscent of an expanding color capacitor: infinitely Lorentz-contracted sheets of $SU(N_c)$ color charge move towards each other (along the  $z$ axis),  pass through each other, and recede.  Color capacitor-like systems have been discussed in the literature in other contexts and we will come back to a comparison later on.  The CGC setup is briefly reviewed in the following.

In the CGC limit nuclei move on the light cone. Their partons can be divided into source partons with large momentum fraction $x$ and classical gluon fields that effectively describe small-$x$ gluons in the nuclear wave functions, as first discussed by McLerran and Venugopalan \cite{McLerran:1993ka,McLerran:1993ni}.
The source is given by a $SU(N_c)$ current $J^\mu = J^{\mu}_{\underline{a}} t^{\underline{a}}$.  Note that we use underlined upper or lower indices for
$SU(N_c)$. The $t^{\underline{a}}$ are the Gell-Mann matrices. We have specified our definitions in appendix \ref{sec:app1}.  The gluon field strength $F^{\mu\nu}$ and its gauge field $A^\mu$ couple to the current through the Yang-Mills equations
\begin{equation}
  \left[ D_\mu, F^{\mu\nu}\right] = J^\nu  \, ,
  \label{eq:ym}
\end{equation}
and the continuity equation
\begin{equation}
  \left[ D_\mu, J^\mu \right] = 0 \, .
  \label{eq:currcons1}
\end{equation}

The internal dynamics of the source are frozen on time scales that describe interactions with probes or other nuclei (the glass in CGC) and are therefore kept fixed on the light cone. In addition, during a collision large angle scatterings of source partons are rare (those would be referred to as hard processes).  The slowing down of source partons through the interaction, in other words the back reaction of the field on the sources, can be significant but at sufficiently large collision energy the source partons are close to the light cone even after the collision. This has been confirmed experimentally even at top RHIC energies where nuclei, represented by the net baryon number carried by the valence quarks, lose about three quarters of their kinetic energy during the collision. Therefore they stay ultrarelativistic throughout 
\cite{Brahms:03}.  This justifies the assumption that a current along the $+$ light cone is invariant, or independent of the $x^+$
coordinate. Because of the practically
infinite Lorentz boost the source is also infinitely thin in the $x^-$ direction, and can therefore be solely described by an $SU(3)$-valued area density 
$\rho(\vec x_\perp)$, where $\vec x_\perp$ is the vector of transverse coordinates.  For our definitions of light cone coordinates we refer the reader to appendix 
\ref{sec:app1}.

We represent two colliding nuclei on the light cone through two currents $J^\mu_{1,2}$ along the $+$ and $-$ light cone, respectively, given by two $SU(3)$ charge densities $\rho_1(\vec x_\perp)$ and $\rho_2(\vec x_\perp)$.  The components of the currents in light cone coordinates are
\begin{align}
  \label{eq:current}
  J^+_1 (x) &= \delta(x^-) \rho_1(\vec x_\perp)\, , \quad J^-_1 (x) = 0 \, ,
  \\
  J^-_2(x) &= \delta(x^+) \rho_2(\vec x_\perp)\, , \quad J^+_2 (x) = 0 \, , \\
  J^i_{1,2} (x) &= 0 \, ,
\end{align}
with $i=1,2$. The total current $J_1^\mu + J_2^\mu$ satisfies the equation of continuity (\ref{eq:currcons1}) if we choose an axial gauge with
\begin{equation}
  x^+ A^- + x^- A^+ = 0.
\end{equation}
We will keep this choice of gauge throughout this section.

We note that nuclei fixed on the light cone will lead to a boost-invariant system after the collision. In particular, the energy-momentum tensor of the gluon field will be boost-invariant. This will be an important caveat when we discuss the global space-time structure of the fireball.  In our calculation global energy, momentum, and angular momentum are \emph{not} conserved as the nuclei are reservoirs for those conserved quantities. In reality, those quantities are finite and conserved. However, we still expect to gain realistic insights of the \emph{rapidity densities} of those quantities as long as we stay far enough away from the final rapidities of the nuclei.  Corrections to the boost invariant approximation can in principle be taken into account \cite{Ozonder:2013moa}.

\subsection{General Shape of the Field}

\begin{figure*}[tb]
\begin{center}
\includegraphics[width=0.45\linewidth]{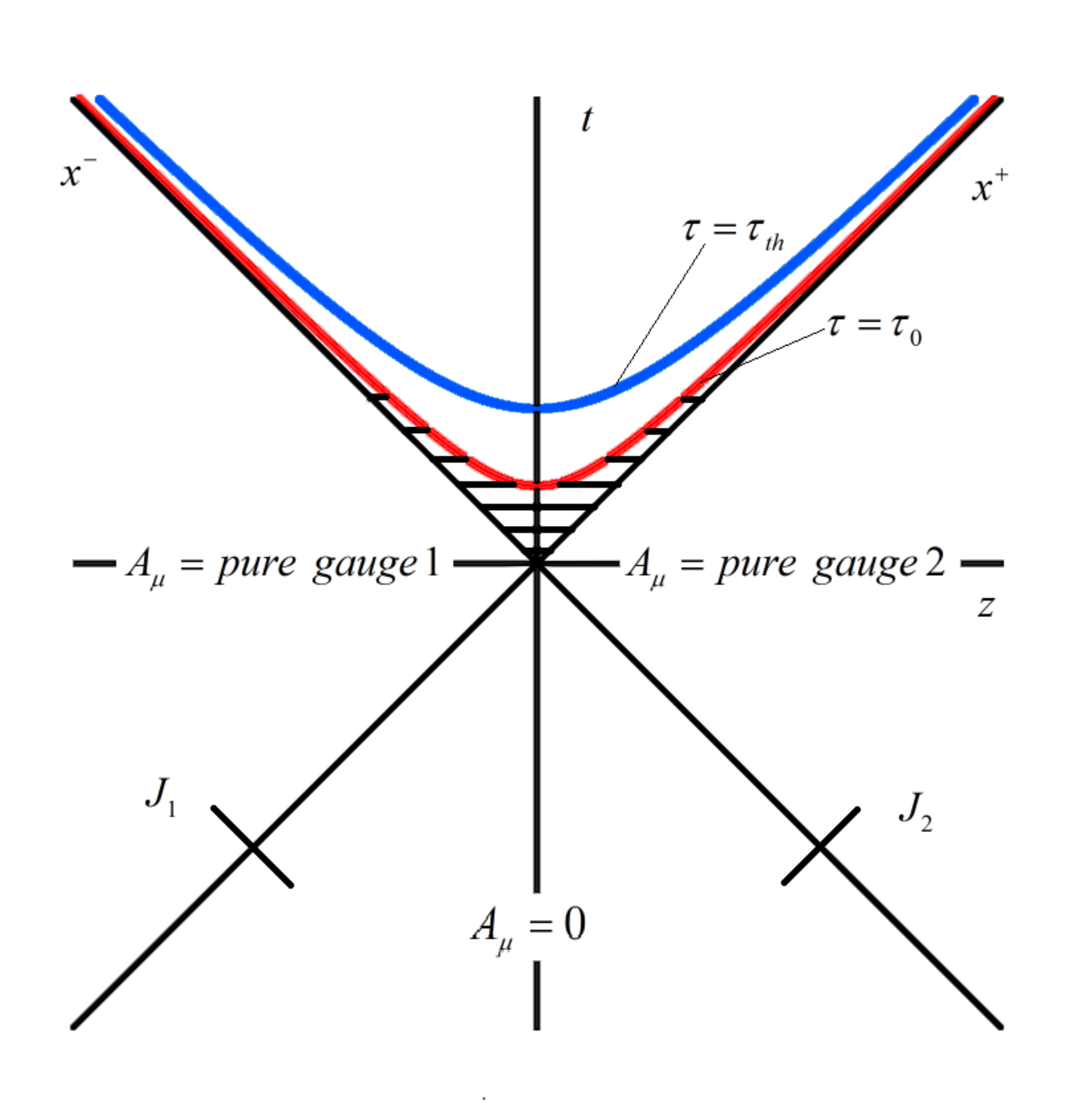}
\caption{(color online) The $z$-$t$-plane with the two currents $J_1$ and $J_2$ given on the $x^+$- and $x^-$-axes, and the four regions given by the solution (\ref{eq:gensol}).  The theoretical limit of the classical approximation in the forward light cone at $\tau=\tau_0$ and the approximate thermalization time $\tau_{\mathrm{th}}$ are shown schematically.}
\label{fig:1}
\end{center}
\end{figure*}

Kovner, McLerran and Weigert were the first to discuss the general space-time structure of the gluon field in the CGC formalism in the collision of two
nuclei \cite{KoMLeWei:95}.  One can write down the following ansatz for the $x^+$-$x^-$-plane:
\begin{align}
  A^+ (x)=& \Theta(x^+)\Theta(x^-) x^+ A(\tau,\vec x_\perp) \, , \nonumber\\
  A^- (x)=& - \Theta(x^+)\Theta(x^-) x^- A(\tau,\vec x_\perp) \, , \nonumber\\
  A^i (x)=& \Theta(x^-)\Theta(-x^+) A_1^i(\vec x_\perp) \label{eq:gensol} \\
         &+ \Theta(x^+)\Theta(-x^-) A_2^i(\vec x_\perp)
           \nonumber\\
         &+ \Theta(x^+)\Theta(x^-) A_\perp^i (\tau,\vec x_\perp) \, ,\nonumber
\end{align}
where again $i=1,2$. The $A_1^i(\vec x_\perp)$ and $A_2^i(\vec x_\perp)$ are the gluon fields of the single nuclei before the collision, which are purely transverse in this gauge. Here $\tau = \sqrt{t^2-z^2}$ is the longitudinal proper time. The $A(\tau,\vec x_\perp)$ and $A_\perp^i(\tau,\vec x_\perp)$ are smooth functions in the forward light cone and describe the field after the collision. They are the glasma fields we will be interested in.  There is no explicit dependence on the space-time rapidity $\eta = {\textstyle{\frac{1}{2}}} \ln \left(x^+/x^-\right)$ in $A$ and $A_\perp^i$, reflecting the boost-invariance of the system. Figure \ref{fig:1} shows the different regions of the light cone including the region of applicability of this work.

In each sector of the light cone the Yang-Mills equations have to be satisfied separately. In the forward light cone they can be written in the convenient form \cite{KoMLeWei:95}
\begin{align}
  & \frac{1}{\tau}\frac{\partial}{\partial\tau}\frac{1}{\tau}
  \frac{\partial}{\partial\tau}
  \tau^2 A  - \left[ D^i ,\left[ D^i,A\right] \right] = 0 \, ,
  \label{eq:eom_noboost1}\\
  & ig\tau \left[ A, \frac{\partial}{\partial\tau}A \right]
  - \frac{1}{\tau} \left[ D^i, \frac{\partial}{\partial\tau}A_\perp^i \right]
  = 0 \, ,
  \label{eq:eom_noboost2}\\
  & \frac{1}{\tau}\frac{\partial}{\partial\tau}
  \tau\frac{\partial}{\partial\tau} A_\perp^i - ig\tau^2 \left[
  A, \left[ D^i,A \right]\right] - \left[ D^j, F^{ji} \right] = 0 \, .
  \label{eq:eom_noboost3}
\end{align}
The field strength tensor in the forward light cone can be expressed in terms of the gauge potentials $A$ and $A_\perp^i$ in this gauge as
\begin{align}
  F^{+-} &= -\frac{1}{\tau} \frac{\partial}{\partial \tau} \tau^2 A, \nonumber
  \\
  F^{i\pm} &= -x^\pm \left( \frac{1}{\tau} \frac{\partial}{\partial\tau}
  A_\perp^i \mp [D^i,A] \right), \label{eq:ffroma} \\
  F^{ij} &= \partial^i A_\perp^j - \partial^j A_\perp^i -ig[A_\perp^i,
  A_\perp^j].  \nonumber
\end{align}
Boundary conditions connect different light cone sectors. The ones for the forward light cone read \cite{KoMLeWei:95}
\begin{align}
  A_\perp^i (\tau=0,\vec x_\perp) &= A_1^i (\vec x_\perp) + A_2^i
  (\vec x_\perp),
  \label{eq:bc_boost1}\\
  A (\tau=0,\vec x_\perp) &= -\frac{ig}{2} \left[ A_1^i
  (\vec x_\perp),A_2^i (\vec x_\perp) \right].
  \label{eq:bc_boost2}
\end{align}
We interpret them as initial conditions for the fields at $\tau=0$ for the fields in the forward light cone $\tau >0$.

Equations\ (\ref{eq:eom_noboost1}) through (\ref{eq:eom_noboost3}) together with the conditions (\ref{eq:bc_boost1}) and (\ref{eq:bc_boost2}) pose the boundary value problem to be solved.  An analytic solution in closed form is not known for the most general case.  The weak field or abelian limit was first treated in \cite{KoMLeWei:95} and will be reproduced below. Several groups have discussed numerical solutions \cite{Krasnitz:2000gz,Lappi:2003bi,Krasnitz:2003jw,Schenke:2012wb}, usually focusing on the plane $\eta=0$.

A different approach to solve the problem was first advocated by some of us in \cite{Fries:2006pv,Fries:2005yc}. The basic idea is as follows. Since the classical approach to CGC loses its applicability very soon after the collision, it will be sufficient to focus on the \emph{near-field}, or small proper times $\tau$. In that case one can utilize a systematic expansion of the Yang-Mills equation in a power series in $\tau$ \cite{Fries:2008vp,Fujii:2008km} . We can expect to find the
leading terms in such an expansion analytically. The natural scale for the convergence of such series should be given by the only time scale in the problem, namely, $1/Q_s$. We will see that this is indeed the case.

\subsection{$\mathbf{\tau}$-Expansion and Recursive Solution}

Let us define the power series
\begin{align}
  A(\tau,\vec x_\perp) &= \sum_{n=0}^\infty \tau^n
  A_{(n)}(\vec x_\perp), \\
  A_\perp^i(\tau,\vec x_\perp) &= \sum_{n=0}^\infty \tau^n
  A_{\perp(n)}^i (\vec x_\perp) \, ,
\end{align}
for the fields parameterizing the gauge potential in the forward light cone.  We devise equivalent power series for the field strength tensor, covariant derivatives and the energy-momentum tensor. We do not include any divergent ($1/\tau^n$) or logarithmic ($\ln \tau$) terms in $\tau$. While the field equations themselves can have divergent solutions they have to be discarded because of the boundary conditions (\ref{eq:bc_boost1}) and (\ref{eq:bc_boost2}). 

We can discuss this point in more detail for the abelianized version of the equations. In the case of weak fields the non-linear terms in the Yang-Mills equations are usually neglected, leading to a greatly simplified abelian version of the boundary value problem. The analytic solution in closed form can be readily found \cite{KoMLeWei:95}.  After applying a Fourier transformation of the transverse coordinate, $\partial^i \to -ik_\perp^i$, Eqs. (\ref{eq:eom_noboost1}) and 
(\ref{eq:eom_noboost3}) take the form of Bessel equations
\begin{align}
  \frac{1}{z} \frac{d^2}{dz^2} {zA} +  \frac{1}{z^2} \frac{d}{dz}
  {zA} + \frac{1}{z} {zA} - \frac{1}{z^3} {zA}  & = 0   \, ,\\
  z^2 \frac{d^2}{dz^2} A^i_\perp +  z \frac{d}{dz} A^i_\perp +
  z^2 A^i_\perp &= 0 \, ,
\end{align}
where $z=k_\perp \tau$. A physical polarization $\nabla^i A^i_\perp =0$ has been chosen for the transverse field.  There are two independent sets of solutions, Bessel functions of the first kind $A \sim J_1(z)/z$, $A_\perp^i \sim J_0(z)$ which are regular at $\tau=0$, and Neumann functions 
$A \sim N_1(z)/z$, $A_\perp^i \sim N_0(z)$ which lead to singular solutions $A \sim z^{-2}$, $A_\perp^i \sim \ln \tau$ for $\tau\to 0$. The solution with Neumann functions is not compatible with Eq.\ (\ref{eq:eom_noboost2}) which imposes $\partial/\partial\tau A^i_\perp = 0$. The singular solution therefore has to be
excluded. 

Let us now return to the solution of the general non-abelian problem.  The power series turns the set of 3 differential equations \eqref{eq:eom_noboost1}, \eqref{eq:eom_noboost2}, and \eqref{eq:eom_noboost3} in $x_\perp$ and $\tau$ into an infinite system of differential equations in $x_\perp$. Amusingly, we can solve this system recursively.  The boundary conditions \eqref{eq:bc_boost1} and \eqref{eq:bc_boost2} provide the starting point of the recursion
\begin{align}
  A_{\perp(0)}^i  &= A_1^i  + A_2^i \, ,
  \label{eq:bc_boost4}\\
  A_{(0)} &= -\frac{ig}{2} \left[ A_1^i , A_2^i \right] \, .
  \label{eq:bc_boost5}
\end{align}
It can be shown that all coefficients of odd powers vanish, $A_{(2k+1)} = 0$ and $A_{\perp(2k+1)}^i = 0$. Finally, one finds the recursion relations for even $n$, $n>1$, to be
\begin{align}
  A_{(n)} =& \frac{1}{n(n+2)} \sum_{k+l+m=n-2} \left[ D^i_{(k)}, \left[
  D^i_{(l)}, A_{(m)} \right] \right] ,  \nonumber \\
  A^i_{\perp(n)} =& \frac{1}{n^2}\left( \sum_{k+l=n-2}
  \left[ D^j_{(k)}, F^{ji}_{(l)} \right] \right. \label{eq:a_recursion2}
  \\
  &+ \left.
  ig \sum_{k+l+m=n-4} \left[ A_{(k)}, [ D^i_{(l)},A_{(m)} ] \right] \right)
  \, .
  \nonumber
\end{align}
One can readily see that these expressions solve (\ref{eq:eom_noboost1}) and (\ref{eq:eom_noboost3}). It is less straight-forward to show that the recursion relation solves Eq.\ (\ref{eq:eom_noboost2}).  One can go order by order in $\tau$, and we have explicitly checked that our recursive solution solves 
Eq. (\ref{eq:eom_noboost2}) up to 4th order in $\tau$.

One can use the abelianized case for a cross check.  After dropping non-linear terms, and after applying a Fourier transformation to the transverse coordinates, the recursive solutions can be easily cast in the form
\begin{align}
  A_{(n)}^{\mathrm{LO}} &= \frac{2}{n!!^2(n+2)} (-k_\perp^2)^{n/2}
  A_{(0)}^{\mathrm{LO}} \, , \>\>\> (n>1)\\
  A^{\mathrm{LO}i}_{\perp(n)} &= \frac{1}{n!!^2} (-k_\perp^2)^{n/2}
  A^{\mathrm{LO}i}_{\perp(0)} \, ,
\end{align}
where the double factorial is $n!! = n(n-2)(n-4)\cdots$ and the index LO signals the abelian case. These terms are just the coefficients of the Bessel functions already discussed above,
\begin{align}
  A^{\mathrm{LO}}(\tau,\vec k_\perp) =&
  \frac{2 A_{(0)}^{\mathrm{LO}}(\mathrm k_\perp)}{k_\perp\tau}
     J_1 \left( k_\perp\tau \right) \, ,
     \\
  A^{\mathrm{LO}i}_{\perp}(\tau,\vec k_\perp) =&
    A^{\mathrm{LO}i}_{\perp(0)}(\mathrm k_\perp) J_0 \left( k_\perp\tau
    \right) \, .
\end{align}
Thus the small-$\tau$ expansion immediately recovers the full abelian solution.

The recursive solution (\ref{eq:a_recursion2}) and its consequences are the basis for the remainder of this manuscript.  A brief discussion on the convergence of the series expansion is in order.  From the abelian case above we infer that the radius of convergence is $\infty$, independent of the charge distributions $\rho_{1,2}$, as long as we are in the weak field limit. In the opposite limit of extremely strong fields one can make the following estimate. Keeping only the maximally
non-abelian terms, we expect from the recursion relations that
\begin{equation}
  |A_{(n)}| \sim |gA|^{n+1} |A|$ \, {\rm and} \, $|A^i_{\perp(n)}| \sim |gA|^n |A| \, ,
\end{equation}
where 
\begin{equation}
  |A| = \sqrt{A^i_1A^i_1} \, ,
\end{equation}
is written in terms of the fields in the initial nuclei before collision. Here we assume head-on collisions of equal nuclei ($\rho_1=\rho_2$) for simplicity of argument.
We anticipate from our results later on that $|g A|^2 \sim g^4 \mu/4\pi$, where $\mu$ is the color charge density of the two incoming nuclei. From the geometric interpretation of the saturation scale $Q_s$ we further have $Q_{s}^2 \sim g^4 \mu$ \cite{Lappi:06}.  Hence we find that parametrically 
\begin{align}
  |A_{(n)}| &\sim Q_s^{(n+1)} |A|  \, , \\
  |A^i_{\perp(n)}| &\sim Q_s^{n} |A| \,  .
\end{align}
This suggests that the convergence radius of the series in this extreme case is indeed parametrically set by the saturation scale, 
$\tau_{\mathrm{conv}} \sim 1/Q_s$.  We can find further phenomenological validation in Sec.\ \ref{sec:homogeneous} when we compare to numerical solutions of the Yang-Mills equations.

\subsection{The Near Field}

\begin{figure*}[tb]
\begin{center}
\includegraphics[width=0.9\linewidth]{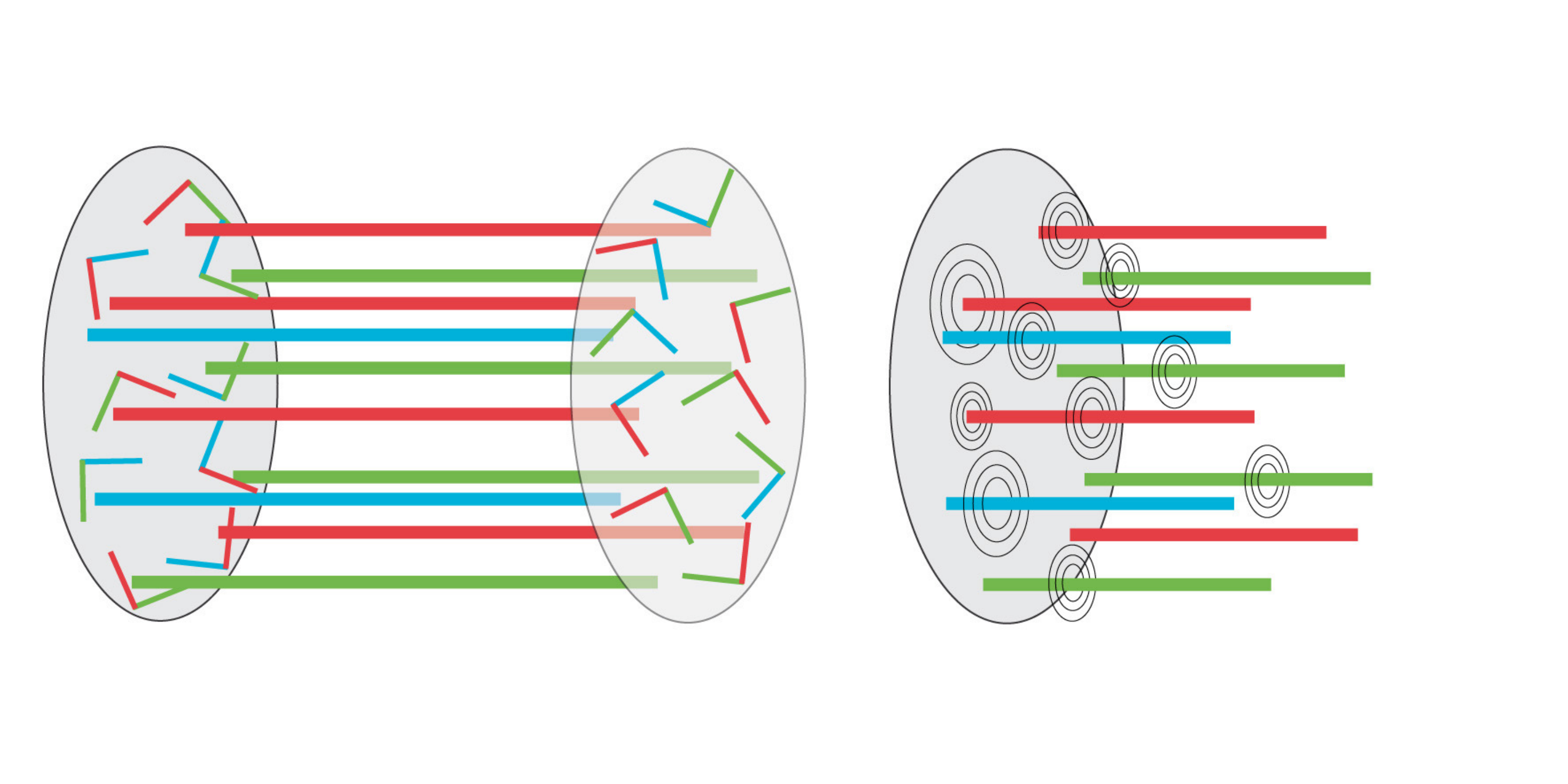}
\caption{(color online) Left: After the collision Lorentz-contracted nuclei with   color charges and transverse fields develop longitudinal fields $E_0$ and $B_0$ between them. Right: Transverse fields between the nuclei are induced by the decrease of longitudinal fields after a short time. Only fields from Faraday's and Ampere's Law are indicated.}
\label{fig:2}
\end{center}
\end{figure*}

A resummation similar to the abelian case seems elusive for the general solution. However, we can analyze the few lowest order terms explicitly.  This amounts to a description of the near field close to the light cone. The series expansions for the gauge potential are
\begin{widetext}
\begin{align}
  A(\tau,x_\perp) &= A_{(0)} + \frac{\tau^2}{8} [D^j, [D^j, A_{(0)} ] ]
  + \frac{\tau^4}{192} [D^k, [D^k, [D^j, [D^j, A_{(0)}]]]] + 
  \frac{ig\tau^4}{48} \epsilon^{ij} [D^i A_{(0)}, D^j B_0] +\mathcal{O}(\tau^6) \, ,\\
  A^i_\perp(\tau, x_\perp) &= A^i_{\perp (0)} + \frac{\tau^2}{4} \epsilon^{ij} [D^j, B_0]
  +\frac{\tau^4}{64} \epsilon^{ij} D^j D^kD^k B_0 - \frac{ig\tau^4}{64} [ B_0, D^i B_0]
  +\frac{ig\tau^4}{16} [A_{(0)}, [D^i, A_{(0)}]]
  + \mathcal{O}(\tau^6)   \, ,
\end{align}
\end{widetext}
where we have used the short hand notation $D^i \equiv D^i_{(0)} = \partial^i - ig A^i_{\perp(0)}$. In the remainder of this work $D^i$ will denote the covariant derivative with respect to the initial gauge field and we will mention explicitly if we refer to covariant derivatives at other times. The $B_0$ is the longitudinal chromo-magnetic field which is discussed below.

Let us carry out an order by order analysis for the field strength tensor
\begin{align}
  \mathbf{E} =& \mathbf{E}_{(0)} + \tau \mathbf{E}_{(1)} + \tau^2
  \mathbf{E}_{(2)} + \ldots \, ,\\
  \mathbf{B} =& \mathbf{B}_{(0)} + \tau \mathbf{B}_{(1)} + \tau^2
  \mathbf{B}_{(2)} + \ldots \, ,
\end{align}
of chromo-electric and chromo-magnetic fields. From here on electric and magnetic always refer to chromo-electric and chromo-magnetic.  The components of the field strength tensor can be readily computed from the gauge potential using Eqs. (\ref{eq:ffroma}).  We observe that only the longitudinal components of the electric and magnetic fields have non-vanishing values at $\tau=0$ \cite{Fries:2005yc}
\begin{align}
  \label{eq:e0}
  E_0 \equiv E^3_{(0)} = F^{+-}_{(0)} &= ig \left[ A_1^i,
  A_2^i \right] \, ,\\
  \label{eq:b0}
  B_0 \equiv B^3_{(0)} = F^{21}_{(0)} &= ig \epsilon^{ij} \left[ A_1^i,
  A_2^j \right] \, .
\end{align}
They can be seen as the seed fields for the glasma developing in the forward light cone.  The transverse fields vanish at $\tau = 0$: $F^{i\pm}_{(0)} = 0$. 

The dominance of longitudinal fields, both electric and magnetic, at early times has been discussed in \cite{Lappi:2006fp,Fries:2006pv} and has since been often relabeled as the occurrence of color flux tubes. They are similar but not directly comparable to QCD strings. QCD strings are a reaction of the QCD vacuum to color charges.  Here we consider fields close to the center of a collision of large nuclei which are far removed from the QCD vacuum. Non-trivial QCD vacuum effects are not included in the classical Yang-Mills picture considered here.  QCD strings have been successfully used to describe collisions of nucleons at large energies \cite{PYTHIA}. It would be desirable to find a natural transition between glasma flux tubes in the center of collisions and QCD strings describing the dynamics at the boundary of the collision zone, but that is beyond the scope of this work.  The initial longitudinal magnetic and electric fields can be of similar strength in the glasma. Figure \ref{fig:2} shows a sketch with nuclei consisting of Lorentz contracted sources and transverse gluon fields, and longitudinal fields stretching between them after the collision.

\begin{figure*}[htb]
\begin{center}
\includegraphics[width=0.9\linewidth]{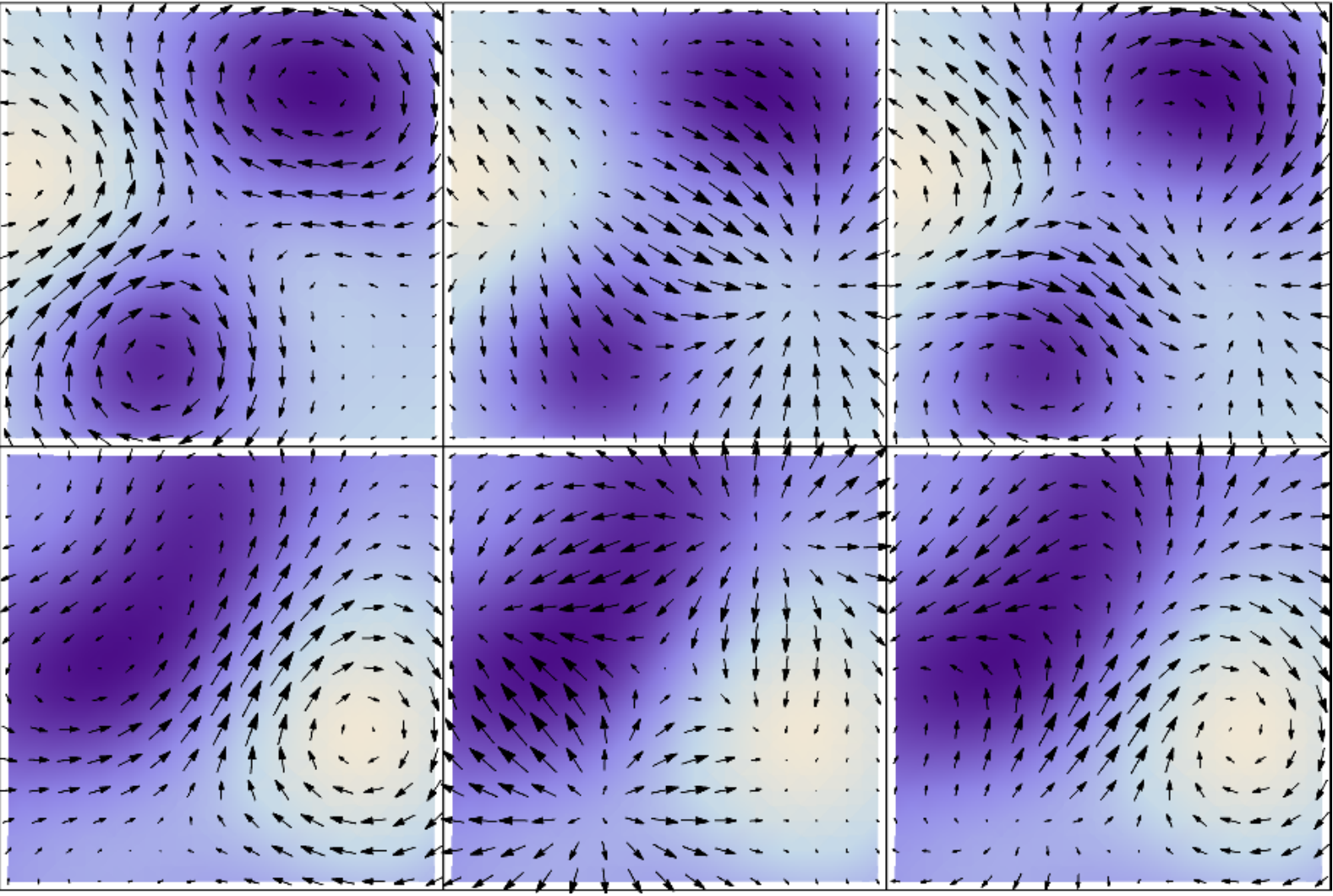}
\caption{(color online) Transverse fields in the $x$-$y$-plane for simple abelian analogue to Eqs.\ (\ref{eq:trfield1}), (\ref{eq:trfield2}) for randomly
  simulated initial fields $A_1^i$ and $A_2^i$. Upper panels: Electric field $E^i$ (arrows) on background of the longitudinal magnetic field $B_0$ (shading).
  Lower panels: Magnetic field $B^i$ on background of $E_0$. Left panels: Rapidity-even terms from Faraday's and Ampere's Law, respectively (corresponding to
  $\eta=0$). Middle panels: Rapidity-odd terms from Gauss' Law.  Right panels: Full transverse fields for $\eta=1$. Length scales of arrows and of $x$ and $y$ axes are arbitrary.  More details can be found in \cite{Chen:2013ksa}.}
\label{fig:fields}
\end{center}
\end{figure*}

It is useful to briefly mention a reinterpretation of the initial longitudinal fields pointed out in \cite{Lappi:2006fp}. This can help to
make connections with some existing phenomenological models in which an exchange of color charge between the nuclei is envisioned at the
time of their overlap \cite{Magas:2000jx,Magas:2002ge}.  Those effective color--anti-color charges on opposite nuclei then lead, in a
quasi-abelian picture, to longitudinal electric fields between the nuclei after they have separated and recede from each other. This appears
similar to the early glasma picture. Note, however, that here the charges $\rho_1$ and $\rho_2$ are strictly kept constant throughout the
collision and the longitudinal field arises from \emph{non-abelian} interactions between the fields of the two nuclei. But to aid our
intuition, we can rewrite the covariant derivatives in Gauss' Laws for chromo-electric and chromo-magnetic fields as ordinary derivatives
and commutator terms which can be interpreted as effective chromo-electric and chromo-magnetic charges $ig[A_1^i,E^i_2]$, $ig[A_2^i,E^i_1]$
and $ig[A_1^i,B^i_2]$, $ig[A_2^i,B^i_1]$, where $E^i_k$ and $B^i_k$ are the transverse fields in nucleus $k$ \cite{Lappi:2006fp}. The
commutators are non-zero when the gauge potential from nucleus 1 can interact with the field of nucleus 2 and vice versa.  Then for $t>0$
the induced charges on opposite nuclei are indeed the negative of each other. Hence, we can also interpret the longitudinal fields as the \emph{abelian} fields generated by additional color charges induced in the collision at $t=0$. 

Going forward in time, we note that the first order in $\tau$ brings no further contribution to the longitudinal fields, $F^{+-}_{(1)}=0=F^{21}_{(1)}$, but it is the
leading order for the transverse fields
\begin{equation}
  F^{i\pm}_{(1)} = -\frac{e^{\pm\eta}}{2\sqrt{2}} \left(
  [ D^j_{(0)},F^{ji}_{(0)} ] \pm [ D^i_{(0)}, F^{+-}_{(0)} ] \right) \, .
\end{equation}
Therefore, the transverse electric and magnetic fields grow linearly from their zero value at $\tau=0$. We can express them in terms of the initial longitudinal fields as \cite{Chen:2013ksa}
\begin{align}
  E^i_{(1)} &= -\frac{1}{2} \left( \sinh\eta [D^i, E_0] + \cosh\eta \,
  \epsilon^{ij}[D^j,B_0] \right) \, ,
  \label{eq:trfield1}  \\
  B^i_{(1)} &= \frac{1}{2} \left( \cosh\eta\, \epsilon^{ij} [D^j, E_0]
  - \sinh\eta [D^i,B_0] \right) \, .
  \label{eq:trfield2}
\end{align}
Recall that we have agreed to the notation $D^i =\partial^i - ig A^i_{\perp(0)}$.  In \cite{Chen:2013ksa} we have discussed extensively how these transverse fields can be understood from the QCD analogues of Faraday's and Amp\`ere's Law for the rapidity even parts and from Gauss' Law for the rapidity-odd components. In particular, it is very natural to expect rapidity-odd transverse fields even in a boost invariant situation.  In Fig. \ref{fig:fields} we show a typical example for the rapidity-even and rapidity-odd initial transverse fields in an abelian example (covariant derivatives are replaced by ordinary derivatives in
(\ref{eq:trfield1}) and (\ref{eq:trfield2})).  Notice how at mid-rapidity field lines are closing around existing longitudinal flux tubes (dark or light colored regions) due to Amp\`ere's and Faraday's Law, while away from mid-rapidity Gauss' Law allows for transverse flux between longitudinal flux tubes.

The first correction to the initial value of the longitudinal fields appears at order $\tau^2$ and in our short notation is
\begin{align}
  E^3_{(2)} &=
  \frac{1}{4} [ D^i , [ D^i , E_0 ]] \, , \\
  B^3_{(2)} &= \frac{1}{4} [ D^i , [ D^i , B_0 ]] \, .
\end{align}
There is no correction to the transverse fields at this order, $F^{i\pm}_{(2)} = 0$.

From order $\tau^3$ on the results become somewhat unwieldy. For this reason we present the expressions for orders $\tau^3$ and $\tau^4$ in Appendix 
\ref{sec:app2}. However, there is no particular reason why one could not in principle go to higher orders in powers of $\tau$.  Generally, the longitudinal fields have only contributions for even powers of $\tau$ and the transverse fields have contributions only for odd powers of $\tau$. 

To summarize this section, we have provided explicit formulas for the initial gluon field to an accuracy
\begin{align}
   E^3 =& E^3_{\mathrm{trunc}} + \mathcal{O}(\tau^6) \, , \\
   E^i =& E^i_{\mathrm{trunc}} + \mathcal{O}(\tau^5) \, ,
\end{align}
for the electric field, and similarly for the magnetic field.

\section{The Energy-Momentum Tensor of the Field}

From the field strength tensor we can easily calculate the energy-momentum
tensor of the field
\begin{equation}
  T^{\mu\nu} = F^{\mu\lambda} F_\lambda^{\nu} + \frac{1}{4} g^{\mu\nu}
  F^{\kappa\lambda} F_{\kappa\lambda} \, .
\end{equation}
For brevity we will often employ a notation where $SU(N_c)$ indices are summed over implicitly unless said otherwise: 
$AB = A^{\underline{a}} B^{\underline{a}} = 2 \tr(AB)$, $\underline{a}=1, \ldots , N_c^2-1$. We will now provide the first few orders in $\tau$ for all components of the energy-momentum tensor
\begin{equation}
  T^{\mu\nu} = T^{\mu\nu}_{(0)} + \tau T^{\mu\nu}_{(1)} + \tau^2 T^{\mu\nu}_{(2)}
  + \ldots
\end{equation}
as functions of the initial longitudinal fields $E_0$ and $B_0$.

\subsection{Initial Energy Density and Pressure}

Only the diagonal elements of $T^{\mu\nu}$ have finite values at $\tau=0$. We define $\varepsilon_0$ to be the initial value for the energy density
\begin{multline}
  \label{eq:eps0}
  \varepsilon_0 = T^{00}_{(0)} = \frac{1}{2} \left( E_0^2 + B_0^2 \right)
  \\ = -\frac{g^2}{2}
  \left( \delta^{ij} \delta^{kl} + \epsilon^{ij}\epsilon^{kl} \right)
  \left([A_1^i,A_2^j] [A_1^k,A_2^l] \right) \, .
\end{multline}
The other diagonal elements of the energy-momentum tensor are
\begin{equation}
  T^{11}_{(0)} = T^{22}_{(0)} =  \varepsilon_0  = -T^{33}_{(0)} \, .
\end{equation}
Hence the structure of the energy-momentum tensor for $\tau\to 0$ is the same as that for a longitudinal field in classical electrodynamics.  There is a maximum pressure anisotropy between the transverse and longitudinal directions. Despite being far from equilibrium we take the liberty to use the notations of longitudinal pressure $p_L = T^{33}$ and transverse pressures $(p_x,p_y) = (T^{11},T^{22})$. We will denote the average transverse pressure as $p_T = (p_x + p_y)/2$.

The initial transverse pressure $p_T =\varepsilon_0$ is large compared to an equilibrated system. A free, relativistic gas with the same energy density would have a transverse pressure $\varepsilon_0/3$.  We expect a comparably large flow of energy due to gradients in the transverse pressure.  The longitudinal pressure 
$p_L = -p_T = -\varepsilon_0$ is equally large and negative. The negative sign is not surprising. Keeping in mind the abelian reinterpretation of the longitudinal field, we expect the opposite sign induced color charges on the nuclei to be attractive. Hence, the initial longitudinal fields would like to decelerate the sources. In fact, this is the mechanism that removes kinetic energy from the nuclei and deposits it as field strength in the space-time region between them.  Here we do not take into account this back reaction of the field on the sources since the nuclei, even at top RHIC energies, seem to stay ultra-relativistic all the time, as discussed before.

The qualitative global behavior of the system then is seemingly easy to predict from the simple form of $T^{\mu\nu}$ at $\tau\to 0$,
\begin{equation}
  T^{\mu\nu}_{(0)} =
  \begin{pmatrix}\varepsilon_0 & 0 & 0 & 0 \\ 0 & \varepsilon_0
    &0 &0 \\ 0& 0& \varepsilon_0 &0 \\ 0& 0& 0& -\varepsilon_0
  \end{pmatrix} \,.
\end{equation}
While the negative longitudinal pressure leads to the deceleration of the colliding nuclei, the transverse pressure forces the system to expand in the transverse direction. This transverse expansion, driven by the pressure of the classical field, is expected to be larger than in an equilibrated relativistic gas \cite{Fries:2005yc,Vredevoogd:2008id}.  We will see that this intuitive picture, while mostly correct, has to have additional features added to it since the energy-momentum tensor above does not have the full information about the classical fields which drive the dynamics.

\begin{figure*}[t]
\begin{center}
\includegraphics[width=0.9\linewidth]{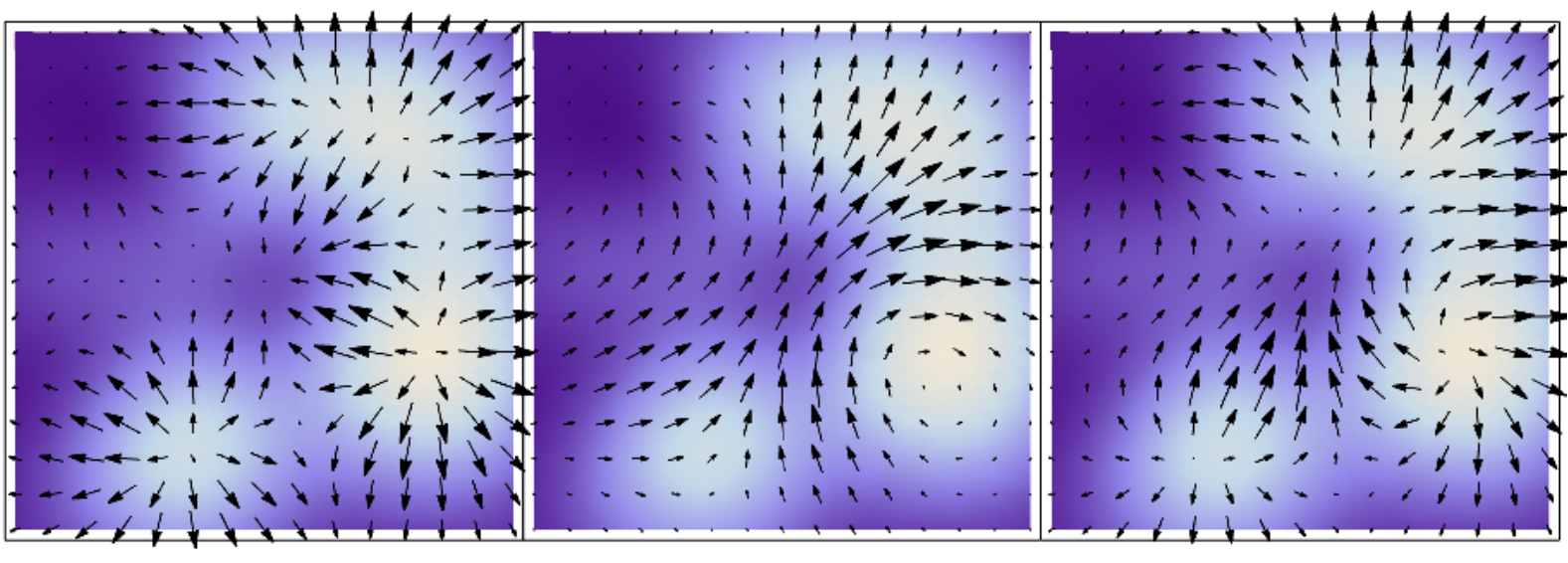}
\caption{(color online) Energy flow in the $x$-$y$-plane in an abelian analogue  for the same random distribution of seed fields $A_1^i$, $A_2^i$ chosen in 
Fig. \ref{fig:fields}. The transverse Poynting vector $T^{0i}$ (arrows)   is drawn with the energy density $\varepsilon_0$ (shading) in the background.  The panels show rapidity-even (left), rapidity-odd (center) and full flow at $\eta=1$ (right). Length scales of arrows and of $x$ and $y$ axes are arbitrary.}
    \label{fig:flow}
\end{center}
\end{figure*}

\subsection{Onset of Transverse Flow}

At the next order, linear in $\tau$, the components $T^{0i}$ and $T^{3i}$, with $i=1,2$, are the only ones to pick up contributions.  They describe the flow of energy and longitudinal momentum into the transverse direction. Note that $T^{0i}$ is the transverse component of the Poynting vector 
$\mathbf{S} = \mathbf{E} \times \mathbf{B}$. Therefore, the transverse expansion expected from the qualitative arguments given above sets in linearly in $\tau$. We have
\begin{align}
  \label{eq:T0i_1}
  T^{0i}_{(1)} =&  \epsilon^{ij} \left( B_0 E^j_{(1)} - E_0 B^j_{(1)} \right)\\
  =& \frac{1}{2} \alpha^i \cosh\eta  +
     \frac{1}{2} \beta^i \sinh\eta  \, , \nonumber \\
  \label{eq:T3i_1}
  T^{3i}_{(1)} =&  - E_0 E^i_{(1)} - B_0 B^i_{(1)} \\
  =& \frac{1}{2} \alpha^i \sinh\eta + \frac{1}{2} \beta^i \cosh\eta \, . \nonumber
\end{align}
We note that we have two contributions to transverse flow. The first term is the flow driven by the gradient of the transverse pressure as we would expect from a hydrodynamic picture \cite{Chen:2013ksa}
\begin{equation}
  \alpha^i = - \nabla^i \varepsilon_0 \, .
\end{equation}
The second term involves the 2-vector
\begin{equation}
  \beta^i = \epsilon^{ij} \left( [D^j,B_0]E_0 - [D^j,E_0]B_0\right) \, .
\end{equation}
The derivation of these and the following expressions is made easier by using a set of $SU(3)$ identities assembled in Appendix \ref{sec:app1}.  These flow terms have first been discussed by some of us in Ref. \cite{Chen:2013ksa}.

The $\beta^i$ defies the naive expectations from our earlier analysis of the initial diagonal energy-momentum tensor.  It is profoundly related to the electric and magnetic fields underlying the energy-momentum tensor.  More precisely, it emerges from the rapidity-odd transverse fields mandated by Gauss' Law.  The $\beta^i$ enhances flow from larger to smaller energy densities in some regions and quenches it in other regions. This can be seen in the example of random abelian fields in Fig. \ref{fig:flow}. This abelian analogue is particularly interesting here since the non-abelian terms in (\ref{eq:T0i_1}) vanish in the event-average as discussed in 
\cite{Chen:2013ksa}. However, they will be important when the field is sampled event-by-event.

The contribution of $\beta^i$ to the energy flow is odd in space-time rapidity $\eta$. We want to stress that its existence does not violate boost-invariance.
Obviously $\beta^i$ will have a role to play when angular momentum and directed flow in the system are studied.

\subsection{Order $\tau^2$: Corrections to Energy Density and Pressure; Longitudinal Flow}

At order $\tau^2$ the diagonal elements of $T^{\mu\nu}$ receive their first corrections and all the previously vanishing components acquire their leading contributions. On the other hand, the transverse flow of energy and longitudinal momentum are not affected,
\begin{equation}
T^{0i}_{(2)} = 0 = T^{3i}_{(2)} \, .
\end{equation}
The expressions for the energy density, the longitudinal flow of energy, and the flow of longitudinal momentum are
\begin{eqnarray}
  T^{00}_{(2)} &=& E_0 E^3_{(2)} + B_0 B^3_{(2)} + \frac{1}{2}
  E^i_{(1)} E^i_{(1)} + \frac{1}{2} B^i_{(1)} B^i_{(1)}   \nonumber \\
    &=& - \frac{1}{4} (\nabla^i \alpha^i + \delta) - \frac{1}{8}  \,
     \nabla^i\beta^i \sinh 2\eta \nonumber \\
    &&+ \frac{1}{8} \, \delta \cosh 2\eta   \, ,
\label{eq:t002}
\end{eqnarray}
\begin{eqnarray}
  T^{03}_{(2)} &=& \epsilon^{ij} E^i_{(1)} B^j_{(1)} \nonumber \\
    &=& - \frac{1}{8} \, \nabla^i\beta^i \cosh 2\eta
    + \frac{1}{8} \delta \sinh 2\eta \, ,
\end{eqnarray}
\begin{eqnarray}
  T^{33}_{(2)} &=& - E_0 E^3_{(2)} - B_0 B^3_{(2)} + \frac{1}{2}
    E^i_{(1)} E^i_{(1)} + \frac{1}{2} B^i_{(1)} B^i_{(1)} \nonumber \\
      &=& \frac{1}{4} (\nabla^i \alpha^i + \delta) - \frac{1}{8}  \,
      \nabla^i\beta^i \sinh 2\eta  \nonumber \\
     && + \frac{1}{8} \delta \cosh 2\eta \, .
\end{eqnarray}
We have used Eqs. (\ref{theta1}) and (\ref{theta2}) to simplify these expressions.  Besides the divergence of the transverse fields, $\alpha^i$ and $\beta^i$, we find a new field that appears in the expressions above, namely
\begin{align}
  \delta =& [D^i, E_0][D^i, E_0]+[D^i, B_0][D^i, B_0] \, .
\end{align}
The divergence of the transverse flow is the expected reaction of the energy density to the initial flow, leading to depletion at the source and accumulation at the sink of the flow field. 

The remaining new contributions to this order give corrections to the transverse pressures
\begin{align}
   T^{ii}_{(2)} =& \frac{(-1)^i}{2} \left( E^1_{(1)}E^1_{(1)} +
    B^1_{(1)}B^1_{(1)} \right. \label{eq:tii} \\ &
     \left. - E^2_{(1)}E^2_{(1)} - B^2_{(1)}B^2_{(1)} \right)
     +  E_0 E^3_{(2)} + B_0 B^3_{(2)}  \nonumber \\
      =& - \frac{1}{4} (- \triangle \epsilon_0 + \delta + (-1)^i \omega ) \, , \nonumber \\
  T^{12}_{(2)} =& - E^1_{(1)} E^2_{(1)} - B^1_{(1)} B^2_{(1)} = \frac{1}{4} \gamma \, .
\end{align}
Here $\triangle$ is the 2-dimensional Laplace operator.  There is no implicit summation over the double index $i =1,2$ in the first equation. The new quantities are
\begin{align}
  \omega =&  \frac{1}{2} \left( [D^1,E_0]^2 - [D^2,E_0]^2 \right. \\
   & \qquad + \left. [D^1,B_0]^2 - [D^2,B_0]^2 \right)  \, , \nonumber \\
  \gamma =& [D^1,E_0][D^2,E_0] + [D^1,B_0][D^2,B_0] \, .
\end{align}
The $\omega$ describes the anisotropy of the pressure in the $x$- and $y$-directions and is therefore responsible for a phenomenon akin to elliptic flow in the transverse plane.

\subsection{Higher Orders}

At order $\tau^3$ the only contributions are the first corrections to the transverse flow $T^{0i}$ and $T^{3i}$. They are
\begin{align}
  T^{0i}_{(3)} =& \epsilon^{ij} \left( B_0 E^j_{(3)} + B^3_{(2)} E^j_{(1)}
  - E_0 B^j_{(3)} - E^3_{(2)} B^j_{(1)} \right) \nonumber \\
  =&  \frac{1}{16} \left( \xi^i \cosh\eta + \zeta^i \sinh\eta  \right) \, ,
   \\
  T^{3i}_{(3)} =& - E_0 E^i_{(3)} - E^3_{(2)}E^i_{(1)} -  B_0 B^i_{(3)}
  - B^3_{(2)} B^i_{(1)} \nonumber \\
   =&  \frac{1}{16} \left( \xi^i \sinh\eta  + \zeta^i \cosh\eta  \right) \, .
  \label{eq:xizeta}
\end{align}
We give the explicit expressions for the flow vectors $\xi^i$ and $\zeta^i$ in Appendix \ref{sec:app2}.

At order $\tau^4$ we have
\begin{align}
T^{00}_{(4)} =& E_0 E^3_{(4)} + B_0 B^3_{(4)} + E^i_{(1)} E^i_{(3)} + B^i_{(1)} B^i_{(3)}
  \nonumber \\ &\quad + \frac{1}{2}
  E^3_{(2)} E^3_{(2)} + \frac{1}{2} B^3_{(2)} B^3_{(2)}   \nonumber \\
  =& \rho + \frac{1}{32} \kappa \cosh 2 \eta   + \frac{1}{32} \sigma \sinh 2 \eta \, ,
\end{align}
\begin{align}
  T^{03}_{(4)} =& \epsilon^{ij} \left( E^i_{(1)} B^j_{(3)} + E^i_{(3)}
  B^j_{(1)}  \right) \nonumber \\
  = & \frac{1}{32}\sigma \cosh 2 \eta  + \frac{1}{32}\kappa \sinh 2 \eta \, ,
\end{align}
\begin{align}
T^{33}_{(4)} =& -E_0 E^3_{(4)} -B_0 B^3_{(4)} + E^i_{(1)} E^i_{(3)} + B^i_{(1)} B^i_{(3)}
  \nonumber \\ &\quad - \frac{1}{2}
  E^3_{(2)} E^3_{(2)} - \frac{1}{2} B^3_{(2)} B^3_{(2)}   \nonumber \\
  =& -\rho + \frac{1}{32}\kappa \cosh 2 \eta   + \frac{1}{32}\sigma  \sinh 2 \eta \, ,
\end{align}
\begin{align}
T^{ii}_{(4)} =& (-1)^i \left( E^1_{(1)}E^1_{(3)} +
    B^1_{(1)}B^1_{(3)} \right. \nonumber \\ & \quad
     \left. - E^2_{(1)}E^2_{(3)} - B^2_{(1)}B^2_{(3)} \right) +  E_0 E^3_{(4)}
     + B_0 B^3_{(4)} \nonumber \\
     & \quad + \frac{1}{2}E^3_{(2)}E^3_{(2)}
      +\frac{1}{2}B^3_{(2)}B^3_{(2)}\nonumber \\
   =& \rho + (-1)^i\lambda \, ,
\end{align}
\begin{align}
T^{12}_{(4)} =& - E^1_{(1)} E^2_{(3)} - B^1_{(1)} B^2_{(3)} -
   E^1_{(3)} E^2_{(1)} - B^1_{(3)} B^2_{(1)} \nonumber \\
  =& \nu \, ,
\end{align}
where the new coefficients $\rho$, $\kappa$, $\sigma$, $\lambda$ and $\nu$ are explicitly given in appendix \ref{sec:app2}.  The expressions for the 
energy-momentum tensor discussed here are accurate up to corrections of order $\tau^5$ for the $T^{0i}$ and $T^{3i}$ components, and up to order
$\tau^6$ for all other components.

\subsection{Checking Energy and Momentum Conservation}

The solutions of the Yang-Mills equations automatically satisfy energy and momentum conservation $\partial_\mu T^{\mu\nu} = 0$.  This can be checked explicitly order by order.  The $\partial_\mu T^{\mu0}$ and $\partial_\mu T^{\mu3}$ receive contributions only for odd powers of $\tau$, whereas
$\partial_\mu T^{\mu i}$ consists only of even powers.  At order $\tau$ we find, for $\nu=0$,
\begin{align}
  \label{eq:econscheck1}
  \partial_\mu T^{\mu0}&\Big|_{\tau} = \left(\cosh\eta \frac{\partial}{\partial \tau}
  - \frac{1}{\tau} \sinh\eta \frac{\partial}{\partial \eta} \right) T^{00}_{(2)}  \nonumber  \\
  &+
  \left(-\sinh\eta \frac{\partial}{\partial \tau}
  + \frac{1}{\tau} \cosh\eta \frac{\partial}{\partial \eta} \right) T^{30}_{(2)} + \nabla^i T^{i0}_{(1)}
   \nonumber \\
  =&-\frac{1}{2}(\nabla^i \alpha^i + \delta) \cosh \eta \nonumber  \\
  &+\frac{1}{2}\cosh \eta \left[ -\nabla^i \beta^i \sinh 2 \eta  + \delta \cosh 2 \eta \right]\nonumber  \\
  &-\frac{1}{2}\sinh \eta \left[-\nabla^i \beta^i \cosh 2 \eta   + \delta \sinh 2 \eta \right]  \nonumber  \\
  &+\frac{1}{2}\nabla^i \alpha^i \cosh \eta +\frac{1}{2} \nabla^i \beta^i \sinh \eta \nonumber  \\
  =&0  \, ,
\end{align}
and similarly for $\nu=3$.

Transverse momentum conservation, $\nu=1,2$, is obvious at zeroth order in $\tau$. From the corresponding equation
\begin{align}
  \partial_\mu T^{\mu i}\Big|_{\tau^0} =&
  \left(\cosh\eta - \sinh\eta \frac{\partial}{\partial\eta}\right)T^{0i}_{(1)}
  \\ &-
  \left( \sinh\eta-\cosh\eta\frac{\partial}{\partial\eta}\right) T^{3i}_{(1)}
  + \nabla^i T^{ii}_{(0)} \nonumber \\ =&
 \alpha^i +  \nabla^i\varepsilon_0
  \, , \nonumber
\end{align}
all terms containing the anomalous flow $\beta^i$ drop out and the remaining expression obviously vanishes using the known result for the hydrodynamic flow $\alpha^i$. Note that the index $i$ is not summed in the term containing $T^{ii}$.

At order $\tau^2$ we have a very similar picture
\begin{eqnarray}
 && \partial_\mu T^{\mu i}\Big|_{\tau^2} =
  \left(3\cosh\eta - \sinh\eta \frac{\partial}{\partial\eta}\right)
  T^{0i}_{(3)}  \nonumber \\
  &&- \left( 3 \sinh\eta-\cosh\eta\frac{\partial}{\partial\eta}\right)
  T^{3i}_{(3)}  \nonumber \\
 && + \nabla^i T^{ii}_{(2)} + \nabla^j T^{ji}_{(2)} \nonumber \\
 && = \textstyle{\frac{1}{4}} \left[ \xi^i - \nabla^i \left( - \triangle \varepsilon_0
  + \delta + (-1)^i\omega  \right) +  \nabla^j \gamma \right]  ,
\end{eqnarray}
with the third order flow contribution $\zeta^i$ dropping out.  Again, the index $i=1,2$ is not summed upon multiple appearance and in addition we define $j$ to be the transverse index with $j \ne i$.  Momentum conservation holds if the equation
\begin{equation}
  \xi^i =  \nabla^i \left( -\triangle \varepsilon_0 + \delta + (-1)^i\omega
  \right) - \nabla^j \gamma  \, ,
  \label{eq:xi}
\end{equation}
is true. It is proven explicitly in Appendix \ref{otau3}.  Similarly, the momentum conservation equations at order $\tau^3$ are
\begin{eqnarray}
64 \rho + 3\kappa + \nabla^i \xi^i=0  \; ,\nonumber \\
3 \sigma + \nabla^i \zeta^i=0 \, .
\end{eqnarray}
We are now confident that we have the correct analytic expressions for the initial gluon field.

\section{Averaging over Color Sources with Transverse Dynamics}

So far we have held the charge distributions $\rho_k$ in the two nuclei fixed. We have expressed the gluon fields and energy-momentum tensor after the collision in terms of the initial longitudinal gluon fields $E_0$ and $B_0$ and the initial transverse gauge potential $A_{\perp (0)}^i$. Those, in turn, are determined by the gauge fields $A_1^i[\rho_1]$ and $A_2^i[\rho_2]$ in the two nuclei before the collision.  In a given nuclear collision the color charge densities $\rho_k$ are not known to us. But if we know the statistical distribution of the densities we could use the results of the last two sections for an 
event-by-event analysis in which color charges $\rho_k$ are statistically sampled according to their distributions. Averages over event samples can then be compared to event averages of experimental data taken.  A CGC event generator of this kind, albeit in 2+1D, has recently been presented in the IP-Glasma framework \cite{Schenke:2012wb}. In that work the time evolution of the gluon fields in the forward light cone was solved numerically.  An event generator based on our results would be able to sample fields or the energy-momentum tensor at early times directly without solving differential equations.  However, in this work we will rather focus on obtaining \emph{analytic} results for the {\it event averaged} energy-momentum tensor. We use the assumptions of the MV model which postulates a simple 
Gaussian distribution of color charges \cite{McLerran:1993ka,McLerran:1993ni}.  We have to generalize the MV model by allowing slowly varying average charge densities in the transverse plane. This will allow us to treat transverse gradients in pressure and their consequences.

We start by observing that the expectation value of the color charge of any nucleus at any given point has to vanish, $\langle \rho(\vec x_\perp) \rangle=0$. However, we expect local fluctuations to occur on typical non-perturbative time scales which are much larger than the nuclear collision time. Hence, the fluctuations are frozen at the moment of the collision. The size of the fluctuations are given by the expectation value $\mu\sim \langle \rho^2(\vec x_\perp) \rangle$ of the squared charge density.  In the MV model it is assumed that fluctuations are Gaussian, uncorrelated in space, and isotropic in $SU(3)$. We will see later that it is necessary to introduce a finite resolution in space to regularize the UV divergence that would emerge from an infinite spatial resolution. Whenever taking averages 
$\langle \ldots \rangle$ we will thus keep in mind that they have to be taken at a finite resolution.  For an observable $O$ measured after the collision of two nuclei the expectation value is given by
\begin{equation}
  \langle O \rangle_{\rho_1,\rho_2} = \int d[\rho_1]d[\rho_2] O(\rho_1,\rho_2)  w(\rho_1) w(\rho_2) \, ,
\end{equation}
where the weight functions $w$ are Gaussians with widths given by the average local charge densities squared, $\mu_1$ and $\mu_2$.

\subsection{The MV Model with Transverse Gradients}
\label{sec:mvtd}

We start with a brief review of the MV model. We implement the averaging over color sources in a given nucleus by fixing the expectation values
\begin{multline}
  \label{eq:chargedensnorm}
  \langle \rho_{\underline{a}}(x^\mp,\vec x_\perp) \rho_{\underline{b}}
  (y^\mp,\vec y_\perp) \rangle = \frac{g^2}{N_c^2-1} \delta_{\underline{ab}}
  \\ \times \lambda( x^\mp,\vec x_\perp) \delta
  (x^\mp - y^\mp ) \delta^2 (\vec x_\perp - \vec y_\perp ) \, ,
\end{multline}
as a precise definition of a (light cone) volume density $\lambda(x^\mp,\vec x_\perp)$ of sources for a nucleus moving along the $+$ or $-$ light cone. In addition, expectation values of any odd number of $\rho$-fields in this nucleus vanish. We have dropped the index $k$ labeling a particular nucleus here for ease of notation, and $\underline{a}$, $\underline{b}$ are explicit $SU(3)$ indices.  We have also made explicit the coupling constant $g$ that was contained in $\rho$ as defined in Eqs.\ (\ref{eq:ym}) and (\ref{eq:current}). The $\lambda$ (and $\mu$) are then volume (and area) number densities of color charge, summed over color degrees of freedom.  Note that the normalization of $\lambda$ and $\mu$ differ by a factor $N_c^2-1$ from many other occurrences in the literature, such as \cite{Lappi:06}.
We allow for a dependence of the expectation value $\lambda$ on both the longitudinal coordinate $x^\mp$ and the transverse coordinate $\vec x_\perp$.

The longitudinal smearing in $x^\mp$ is necessary to compute expectation values correctly, as first realized in  \cite{JMKMW:96}. A nucleus must be given a small, but finite, thickness across the light cone which we will do by introducing
\begin{equation}
  \label{eq:hfac}
  \lambda(x^\mp,\vec x_\perp) = \mu(\vec x_\perp)  h(x^\mp) \,.
\end{equation}
Here $h$ is a non-negative function with finite width around $x^\pm = 0$ and normalized such that 
\begin{equation}
  \int dx^\mp \lambda(x^\mp,\vec x_\perp) = \mu(\vec x_\perp) \, .
\end{equation}
It is not necessary to specify the shape of $h$ further.

We have introduced the dependence of the charge densities $\lambda$ and $\mu$ on $\vec x_\perp$  as a generalization of the original MV model, where the nuclei are assumed to be infinitely large in the transverse direction and on average invariant under rotations and translations. Real nuclei break these symmetries; in order to generate a non-trivial transverse dynamics we need to investigate how the results in the MV model generalize when small deviations from these symmetries are
allowed.  Our guiding principle is that, on transverse length scales that are equal to or smaller than the scale of color glass, $1/Q_s$, the gluon field is described by the well-defined color glass formalism. On larger length scales other dynamical effects, for example from the nucleonic structure of the nucleus, appear and can be parameterized by the dependence of $\mu$ on $x_\perp$.  Here we introduce an infrared length scale $1/m$.  We must require that $\mu$ varies by a negligible amount on length scales smaller than $1/m$.  Explicitly we require that
\begin{multline}
  \left| \mu (\vec x_\perp) \right| \gg
  m^{-1} \left|\nabla^i \mu (\vec x_\perp) \right| \\
 \gg
  m^{-2} \left| \nabla^i \nabla^j \mu (\vec x_\perp)\right|
  \gg \ldots \, .
  \label{eq:gradientcondition}
\end{multline}
Then $m$ is an infrared energy scale which separates color glass physics from long wavelength dynamics.  It is necessary to have the hierarchy 
\begin{equation}
  1/Q_s \ll 1/m \ll R_A \, ,
\end{equation}
where $R_A$ is the nuclear radius.

We have two main goals in this extended MV model: (i) Observables must be well behaved under small deviations from translational and rotational invariance, otherwise the original MV model would not be infrared safe. In practice this means that observables should be only weakly dependent on the infrared scale. We  will explicitly check this condition below.  (ii) The results will allow simple long-wavelength dynamics, expressed in an expansion in gradients of $\mu$, which is compatible with color glass physics at small distances.  In practice this will allow us to safely apply the MV model locally to realistic nuclei as long as the location is sufficiently far away from the surface of the nucleus where the density $\mu$ starts to fall off quickly.

\subsection{The Gluon Distribution}
\label{sec:aa}

The most important expectation value of fields in a single nucleus is the two-point function $\langle A(\vec x_\perp) A(\vec y_\perp)\rangle$ which, in light cone gauge, is related to the gluon distribution.  The Yang-Mills equations (\ref{eq:ym}) for a single nucleus on the $+$ light cone are most easily solved in a covariant gauge first where $A_{\mathrm{cov}}^\mu = \delta^{\mu+} \alpha$. The equations reduce to
\begin{equation}
  \Delta \alpha (x^-, \vec x_\perp) = - \rho_{\mathrm{cov}}(x^-, \vec x_\perp) \, ,
\end{equation}
where the Laplace operator $\Delta$ acts on the transverse coordinates.  The explicit solution is
\begin{equation}
  \alpha (x^-, \vec x_\perp) = \int dz^2_\perp G(\vec x_\perp - z_\perp)
  \rho_{\mathrm{cov}}(x^-, \vec z_\perp) \, ,
\end{equation}
with a Green's function $G(x_\perp) = -\ln(x_\perp^2/\Lambda^2)/4\pi$ where $\Lambda$ is an arbitrary length scale. However, we will be better served by introducing a physically motivated regularization through a gluon mass $m$ which can be inserted into the Fourier transformation of the Green's function 
$\tilde G (k) = 1/k^2 \to 1/(k^2+m^2)$  \cite{Fujii:2008km}.  This gluon mass could be an unrelated infrared scale, but for simplicity we will choose it to be the same as the IR cutoff in the gradient expansion of $\mu$ introduced in the previous subsection.  Including the gluon mass leads to the Green's function
\begin{equation}
  G(x_\perp) = \frac{1}{2\pi} K_0(m x_\perp) \, ,
  \label{eq:k0}
\end{equation}
where $K_0$ is a modified Bessel functions. This Green's function reproduces the previous expression in the limit $m\to 0$ with $\Lambda = 2e^{-\gamma_E}/m$, where $\gamma_E$ is Euler's constant.

The two-gluon correlation function in covariant gauge can then be easily derived from (\ref{eq:chargedensnorm}) as
\begin{multline}
  \label{eq:covcorr}
  \langle \alpha_{\underline{a}}(x^-,\vec x_\perp) \alpha_{\underline{b}}
  (y^-,\vec y_\perp) \rangle = \frac{g^2}{N_c^2-1} \delta_{\underline{ab}}
  \\ \times
  \delta(x^- - y^-) \gamma(x^-,\vec x_\perp, \vec y_\perp) \, .
\end{multline}
Here we have introduced another Green's function
\begin{multline}
  \label{eq:gammadef}
  \gamma(x^-,\vec x_\perp,\vec y_\perp) \\  =
  \int d^2 \vec z_\perp G(\vec x_\perp - \vec z_\perp) G(\vec y_\perp - \vec z_\perp)
  \lambda(x^-, \vec z_\perp) \, .
\end{multline}
We will see that $\gamma$ depends strongly on the IR regularization scale $m$. In the limit $r=|\vec y-\vec x| \to 0$ it diverges like $1/m^2$; 
cancellation of this divergence for observables is a critical test of the theory.

The gluon field $A^i$ in light cone gauge can be derived from the gluon field in covariant gauge with the help of the Wilson line
\begin{equation}
  U(x^-,x_\perp) = \mathcal{P} \exp\left[ -ig \int_{-\infty}^{x^-}
  \alpha(z^-,\vec x_\perp) dz^- \, \right].
\end{equation}
Here $\mathcal{P}$ denotes path ordering of the fields $\alpha$ from right to left. One can show that the correct gauge transformation to arrive at the light cone gauge potential is \cite{JMKMW:96}
\begin{equation}
  A^j(x^-,\vec x_\perp) = \frac{i}{g} U(x^-,\vec x_\perp) \partial^j
  U^\dagger(x^-,\vec x_\perp)  \, .
\end{equation}
We apply this gauge transformation to the field strength tensors in covariant gauge to obtain the corresponding tensors in light cone gauge, 
$F = UF_{\mathrm{cov}}U^\dagger$. Their correlation function is 
\begin{multline}
  \label{eq:FFcorr}
  \langle F^{+i}_{\underline{a}}(x^-,\vec x_\perp)F^{+j}_{\underline{b}}
  (y^-, \vec y_\perp) \rangle \\ =
  \left\langle \left(\mathcal{U}^\dagger_{\underline{ac}} \partial^i
  \alpha_{\underline{c}} \right) (x^-,\vec x_\perp)
  \left(\mathcal{U}^\dagger_{\underline{bd}} \partial^j
  \alpha_{\underline{d}} \right) (y^-,\vec y_\perp) \right\rangle  \, .
\end{multline}
In the above expression we have expressed the Wilson lines $U$ by their counterparts in the adjoint representation,
$\mathcal{U}$, by virtue of the relation
\begin{equation}
  Ut_{\underline{a}}U^\dagger = \mathcal{U}_{\underline{ab}} t_{\underline{b}} \, .
\end{equation}

Let us take a small detour to discuss expectation values of adjoint, parallel Wilson lines in the MV model \cite{JMKMW:96}. A systematic study was carried out by Fukushima and Hidaka \cite{Fukushima:2007dy}.  For a single line we obtain
\begin{multline}
  \label{eq:ucorr}
  \langle \mathcal{U}_{\underline{ab}} (x^-, \vec x_\perp) \rangle
  = \delta_{\underline{ab}}
  \exp \bigg[ -\frac{g^4N_c}{2(N_c^2-1)}  \\ \times
  \int_{-\infty}^{x^-} \gamma(z^-,\vec x_\perp,
  \vec x_\perp) dz^- \bigg] .
\end{multline}
This expectation value is suppressed since $\gamma(z^-,\vec x_\perp,\vec y_\perp)$ tends to diverge in the limit $m \to 0$.  For a double line we have
\begin{equation}
  \label{eq:uucorr}
  \left\langle \mathcal{U}_{\underline{ab}} (x^-,\vec x_\perp)
  \mathcal{U}_{\underline{cd}} (x^-,\vec y_\perp) \right\rangle
   = \delta_{\underline{ad}}\delta_{\underline{bc}}
   d(x^-,\vec x_\perp,\vec y_\perp) \, ,
\end{equation}
where 
\begin{multline}
  d(x^-,\vec x_\perp,\vec y_\perp) = \exp\left[ \frac{g^4 N_c}{2(N_c^2-1)} \right. \\ \times
  \left. \int_{-\infty}^{x^-} dz^-
  \Gamma(z^-, \vec x_\perp, \vec y_\perp)  \right]  \, ,
\end{multline}
is the exponentiation of the integral of
\begin{multline}
  \Gamma(z^-, \vec x_\perp, \vec y_\perp) =
  2\gamma(z^-,\vec x_\perp,\vec y_\perp) \\
  - \gamma(z^-,\vec x_\perp,\vec x_\perp) - \gamma(z^-, \vec y_\perp, \vec y_\perp)
  \label{eq:cancel}
\end{multline}
along the light cone.  This $\Gamma$ is a subtracted version of $\gamma$.  In the original MV model the subtraction in $\Gamma$ removes the $1/m^2$ 
singularity in $\gamma$ for small $m$ and renders the exponential $d$ finite. In particular, $\Gamma(x^-,\vec x_\perp,\vec y_\perp)$
vanishes in the ultraviolet limit $\vec y_\perp \to \vec x_\perp$. We will show below that this crucial cancellation is still valid for our generalization.  Here we have dropped contributions from non-color singlet pairs as in \cite{Fukushima:2007dy}.

Now we return to the discussion of the correlation function of fields.  One can prove that the only possible contraction of fields on the right hand side of 
Eq. (\ref{eq:FFcorr}) comes from the factorization of expectation values 
$\langle \mathcal{U}^\dagger\mathcal{U}^\dagger \rangle \langle \partial^i \alpha \partial^j \alpha \rangle$ \cite{FillionGourdeau:2008ij}.  The second factor can be determined from 
Eq. (\ref{eq:covcorr}) as
\begin{multline}
  \langle \partial^i \alpha_{\underline{a}}(x^-,\vec x_\perp) \partial^j
  \alpha_{\underline{b}} (y^-,\vec y_\perp) \rangle = \frac{g^2}{N_c^2-1}
  \delta_{\underline{ab}} \\ \times
  \delta(x^- - y^-) \nabla^i_x \nabla^j_y
  \gamma(x^-,\vec x_\perp , \vec y_\perp) \, .
\end{multline}
Together with Eq.\ (\ref{eq:uucorr}) this leads to the result
\begin{multline}
  \label{eq:FFgd}
  \langle F^{+i}_{\underline{a}}(x^-,\vec x_\perp)F^{+j}_{\underline{b}}
  (y^-, \vec y_\perp) \rangle = \frac{g^2}{N_c^2-1} \delta_{\underline{ab}}
  \delta(x^- - y^-)  \\ \times
  \left[ \nabla^i_x \nabla^j_y
  \gamma(x^-,\vec x_\perp ,\vec y_\perp) \right]
  d(x^-,\vec x_\perp,\vec y_\perp) \, ,
\end{multline}
for the expectation value of fields in light cone gauge.  The correlation function of two gauge potentials in light cone gauge follows from an integration with retarded boundary conditions
\begin{equation}
  A^i(x^-,\vec x_\perp) = -\int_{-\infty}^{x^-} dz^- F^{+i} (z^-,\vec x_\perp) \, .
\end{equation}
One integral is easily evaluated to give
\begin{multline}
  \langle A^i_{\underline{a}} (x^-, \vec x_\perp)  A^j_{\underline{b}}
  (y^-, \vec y_\perp) \rangle =  g^2  \delta_{\underline{ab}}
  \frac{2 \nabla^i_x \nabla^j_y \gamma(\vec x_\perp , \vec y_\perp)}{g^4 N_c
    \Gamma(\vec x_\perp,\vec y_\perp)} \\ \times
  \int_{-\infty}^{\min\{x^-,y^-\}}   d x'^- \frac{\partial}{\partial x'^-}
  \exp\bigg[ \frac{g^4 N_c}{2(N_c^2-1)}
    \\  \times  \Gamma(\vec x_\perp,\vec y_\perp)
  \int_{-\infty}^{x'^-} dz^- h(z^-) \bigg] \, .
\end{multline}
Note that we have used Eq. (\ref{eq:hfac}) which allows us to factor $h(x^-)$ from $\gamma(\vec x_\perp , \vec y_\perp)$ and 
$\Gamma(\vec x_\perp , \vec y_\perp)$. We have formally defined $\gamma (\vec x_\perp,\vec y_\perp)$ as the integral of  
$\gamma (x^-, \vec x_\perp,\vec y_\perp)$ over $x^-$ from $-\infty$ to $+\infty$, and similarly for $\Gamma$.  Here we have rewritten one factor of $h(x^-)$ 
as a derivative $\partial/\partial x'^-$ of the exponential.
 
We can now evaluate the second integral. We will only be interested in  $\min\{x^-,y^-\} > 0$.  Upon taking the limit of vanishing width of $h$ we find that the fields are independent of the coordinates $x^-$ and $y^-$ as long as $\min\{x^-,y^-\} > 0$. We simply write
\begin{multline}
  \label{eq:AA}
  \langle A^i_{\underline{a}} (\vec x_\perp)  A^j_{\underline{b}}
  (\vec y_\perp) \rangle =  2g^2  \delta_{\underline{ab}}
  \frac{\nabla^i_x \nabla^j_y \gamma(\vec x_\perp , \vec y_\perp)}{g^4 N_c
    \Gamma(\vec x_\perp,\vec y_\perp)}  \\ \times
  \left( \exp\left[ \frac{g^4 N_c}{2(N_c^2-1)}\Gamma(\vec x_\perp,\vec y_\perp)
    \right] -1 \right) \, .
\end{multline}
This result holds for both the MV model \cite{JMKMW:96} and our generalization of it.
 
Before proceeding, let us write down the correlation function of two gluon fields when we formally take the ultraviolet limit $\vec y_\perp \to \vec x_\perp$. In that limit $\Gamma \to 0$, and we can expand the exponential function around 0, using only the two leading terms, to arrive at the simpler expression
\begin{multline}
  \label{eq:AA2}
  \langle A^i_{\underline{a}} (\vec x_\perp)  A^j_{\underline{b}}
  (\vec x_\perp) \rangle
  \\ = \delta_{\underline{ab}}  \frac{g^2}{N_c^2-1}
  \nabla^i_x \nabla^j_y \gamma(\vec x_\perp , \vec y_\perp) \Big|_{\vec y_\perp \to
    \vec x_\perp}
  \, .
\end{multline}
For further evaluation of the gluon distribution we have to understand the correlation functions $\gamma$ and $\Gamma$. 

\subsection{Gluon Fields in the MV Model with Transverse Gradients}

The cancellation of the singularity in $\gamma$ through the subtraction in Eq.\ (\ref{eq:cancel}) is a classic result of the original MV model for constant (in transverse coordinates) average charge densities. We will now show that this result holds for the inhomogeneous charge densities $\lambda$ that we have permitted.  More precisely, we will show how expectation values of fields, like the gluon distribution above, can be systematically expanded in gradients of $\mu$.  Let us introduce center and relative coordinates for two points $\vec x_\perp$ and $\vec y_\perp$ in the transverse plane via $\vec R = (\vec x_\perp+\vec y_\perp)/2$ and
$\vec r = \vec y_\perp-\vec x_\perp$.  The discussion in this subsection will use the area charge density $\mu$, but all results apply in a straightforward way to the generalized density $\lambda$ and correlation functions not integrated over $x^-$.

In the original MV model with constant $\mu(\vec x_\perp) =\mu_0$, we can easily calculate the correlation function $\gamma$ defined in Eq. (\ref{eq:gammadef}) to be
\begin{eqnarray}
  \gamma_0(r) &\equiv& \gamma_0(\vec x_\perp,\vec y_\perp) \nonumber \\
&=&  \mu_0 \int d^2 z_\perp G(\vec x_\perp -\vec z_\perp) G(\vec y_\perp  -\vec z_\perp) \nonumber \\ 
&=& \mu_0 \int \frac{d^2 k_\perp}{(2\pi)^2} e^{i\vec k_\perp \vec r}
  \frac{1}{(k_\perp^2+m^2)^2}  \nonumber \\ 
&=& \mu_0 \frac{r}{4\pi m}K_1(mr) \, ,
  \label{eq:k1}
\end{eqnarray}
where $m$ is the same gluon mass introduced as a IR regulator before.  The $\gamma_0$ only depends on the relative distance $r=|\vec r|$ due to isotropy and translational invariance.  As mentioned before, $\gamma_0$ exhibits a quadratic dependence on the infrared cutoff $m$ for small $r$, specifically it is 
$\gamma_0(0) = \mu_0/4\pi m^2$.

On the other hand, this singularity cancels in the subtracted 2-point function (\ref{eq:cancel}). In the UV limit $r \to 0$ the leading term is
\begin{multline}
  \Gamma_0(r) = 2\gamma_0(r) - 2\gamma_0(0)
  \\ = \mu_0 \frac{r^2}{8\pi} \left(\ln \frac{r^2m^2}{4} +2\gamma_E-1\right)
  + \mathcal{O}(m^2 r^4)  \, .
\end{multline}
This is the equivalent of the result in \cite{JMKMW:96} using a finite gluon mass regularization. The $\Gamma_0$ only exhibits a weak logarithmic dependence on $m$ for small $r$.

Let us now check that the same cancellation takes place if $\lambda$ is weakly varying on length scales $1/m$ as permitted here.  We are only interested in typical values of $r = |\vec y_\perp-\vec x_\perp| \lesssim Q_s^{-1} \ll m^{-1}$ since we will later take the UV limit.  We recall from Eq.\ (\ref{eq:k0}) that the Green functions 
$G(z_\perp)\sim K_0(mz_\perp)$ fall off on a scale $1/m \gg r$.  With this clear separation of length scales we can restrict ourselves to the first few terms of a Taylor expansion of $\mu$ around $\vec R$ in the calculation of $\gamma$
\begin{equation}
  \mu(\vec z_\perp) = \mu(\vec R) + (\vec z_\perp-\vec
  R)^i \nabla^i \mu(\vec R) + \ldots \,.
\end{equation}
This leads to
\begin{multline}
  \label{eq:gammaexpand}
  \gamma(\vec R,\vec r) \equiv \gamma(\vec x_\perp,\vec y_\perp)  \\
  = \gamma_0(\vec R,r) + \frac{1}{2} \nabla^i\nabla^j \mu(\vec R)
  \gamma^{ij}(\vec r) + \ldots \, .
\end{multline}
Here we have $\gamma_0(\vec R,r) = \mu(\vec R) r K_1(mr) /4\pi m$ analogous to Eq. (\ref{eq:k1}), representing the constant term. The linear term vanishes because
\begin{equation}
  \int d^2 \vec z_\perp G(\vec z_\perp + \vec r/2) G(\vec z_\perp - \vec
  r/2) z_\perp^i = 0 \, .
\end{equation}
The second order term is
\begin{eqnarray}
\!\!\!\!\!\!\!\!\!\!  \gamma^{ij} &=& \int d^2 \vec z_\perp G(\vec z_\perp) G(\vec r-\vec z_\perp)
  z_\perp^i z_\perp^j \nonumber  \\
  &=& \delta^{ij} \frac{r^2}{24\pi m^2} K_2(mr) +
  \frac{r^ir^j}{r^2} \frac{r^3}{48 \pi m} K_1(mr) \, .
\end{eqnarray}
These correlations functions can be conveniently computed in Fourier space, similar to the technique in Eq. (\ref{eq:k1}).

The subtraction of $\gamma(0)$ removes the leading quadratically divergent term in $m$ as in the original MV model. We can expand $\gamma_0$ and $\gamma^{ij}$ for small $mr$. For $\Gamma$ this leads to
\begin{multline}
  \label{eq:ggamma}
  \Gamma(\vec R,\vec r) = \mu(\vec R) \frac{r^2}{8\pi}
  \left(\ln\hat m^2 r^2 -2\right) + \mathcal{O} (\mu m^2 r^4)\\
  + \nabla^i\nabla^j \ \mu(\vec R) \left[ -\delta^{ij} + 
     \frac{r^i r^j}{r^2} \right] \frac{r^2}{48\pi m^2} \\
   + \mathcal{O}([\nabla^2 \mu] m^0r^4) + \mathcal{O}(\nabla^4\mu)  \, ,
\end{multline}
where $\hat m = m \exp(\gamma_E+1/2)/2 \approx 1.47 m$.  Indeed, the dependence on the cutoff $m$ is at most logarithmic for the small variations of $\mu$ that are permitted. Even though we could take the expansion (\ref{eq:gammaexpand}) farther we will never keep gradients of $\mu$ larger than second order. Higher derivatives will be hard to control phenomenologically, and it is now obvious that condition (\ref{eq:gradientcondition}) guarantees that the derivative correction
in our result for $\Gamma$ is small.

Besides the subtracted correlation function $\Gamma$ we need the double derivative $\nabla^i_x\nabla^j_y \gamma(\vec x,\vec y)$ for the gluon distribution (\ref{eq:AA2}). As discussed above, we neglect gradients of $\mu$ beyond second order. We have two mass scales in the problem, $m \ll Q$, which could cancel the dimensions of energy$^{-1}$ introduced by the gradient expansion.  The UV cutoff $Q$ was introduced earlier as the resolution scale in the transverse plane. We anticipate that in the next step of the calculation we take the limit $r\to 0$, meaning that explicit factors of $r$ will turn into powers of $1/Q$. We only keep terms like 
$m^{-1}\nabla \gg Q^{-1}\nabla \sim r\nabla$. In other words, we drop terms that are suppressed by additional powers of the large scale $Q$.  Thus we arrive at
\begin{multline}
  \label{eq:ddgamma}
  \nabla^i_x\nabla^j_y \gamma(\vec x,\vec y) \\ = \mu(\vec R) \frac{1}{4\pi}
  \left[ \delta^{ij} K_0(mr) - \frac{r^ir^j}{r^2} mr K_1(mr) \right] \\
  + \left[ 2 \nabla^i \nabla^j \mu (\vec R)  + \triangle
  \mu(\vec R) \delta^{ij} \right] \frac{mr}{48\pi m^2} K_1(mr) \\
  + \mathcal{O}(\nabla^3 \mu, r^2 \nabla^2 \mu, \ldots)  \, ,
\end{multline}
where any gradients $\nabla^i$ on the right hand side act only on $\mu(\vec R)$.  Note that terms with single derivatives $\nabla^i \mu$ are power suppressed.
Now we take the formal limit $r \to 0$. No dependence on the direction of $\vec r$ should remain in this limit and we keep only terms isotropic in $\vec r$ by setting $r^i r^j /r^2 \to \delta^{ij}/2$, The leading terms of the correlation function with two derivatives in the ultraviolet limit are
\begin{multline}
   \nabla^i_x\nabla^j_y \gamma(\vec R,\vec r)\Big|_{r \to 0}
  = - \mu(\vec R) \frac{1}{8\pi} \delta^{ij}  \ln (\hat m^2 r^2)
  \\ + \frac{1}{48 \pi m^2} \left[ 2 \nabla^i \nabla^j \mu (\vec R) + 
    \triangle  \mu (\vec R) \delta^{ij}  \right]  \, .
\label{eq:ddgamma2}
\end{multline}
Equations (\ref{eq:ggamma}) and (\ref{eq:ddgamma2}), together with Eq. (\ref{eq:AA}) without the gradient corrections, reproduce the standard result for
the 2-point function in the MV model \cite{JMKMW:96,Lappi:06}
\begin{equation}
  \langle A^i_{\underline a}(\vec x_\perp) A^i_{\underline a}(\vec
  y_\perp)\rangle = \frac{4(N_c^2-1)}{g^2 N_c r^2} \left( 1-
  (\hat m^2 r^2)^{\frac{g^4 N_c}{16 \pi(N_c^2-1)} \mu r^2} \right) .
\end{equation}
Remember that our definition of $\mu$ has an additional factor $N_c^2-1$ compared to Refs.\ \cite{JMKMW:96,Lappi:06}.

Here we are strictly interested in the UV limit $r\to 0$ regularized by a resolution length scale $1/Q$. Plugging (\ref{eq:ddgamma2}) directly into (\ref{eq:AA2}) we obtain
\begin{multline}
  \label{eq:AA3}
  \langle A^i_{\underline{a}} (\vec x_\perp)  A^j_{\underline{b}}
  (\vec x_\perp) \rangle
  = \delta_{\underline{ab}}  \frac{g^2 \mu (\vec x)}{8\pi (N_c^2-1)}
   \left[ \delta ^{ij} \ln \frac{Q^2}{\hat m^2}  \right. \\  + \left.
   \frac{ \nabla^k \nabla^l \mu(\vec x)}{6m^2 \mu(\vec x)}
   \left( \delta^{kl} \delta^{ij} +2 \delta^{ik}
   \delta^{jl} \right) \right]   \, ,
\end{multline}
keeping all leading terms in powers of $1/Q$ up to second order in gradients.  We have made the replacement $r \to 1/Q$ in the logarithm,
which is equivalent to imposing $Q$ as the momentum cutoff in a Fourier representation. The typical transverse momentum of gluons in the nuclear wave function is
given by the saturation scale $Q_s$. Here we can take 
$Q_s^2 \sim g^4 \mu/(N_c^2-1)$ in accordance with \cite{Lappi:06} 
(accounting for the factor $N_c^2-1$ difference in the definition of $\mu$). 
$Q_s$ is the largest scale in the problem and thus the ultraviolet scale $Q$ 
for a single nucleus should be proportional to $Q_s$ with some numerical 
factor, $Q=KQ_s$.

\subsection{Higher Twist Gluon Correlation Functions}
\label{hightwist}

For the components of the energy-momentum tensor beyond the leading term in the $\tau$-expansion, we will need expectation values of gluon fields beyond the 
2-point function. We will compute those correlation functions in this subsection. With more fields or more derivatives these are akin to higher twist distributions of the gluon field. The power counting technique in $1/Q$ we introduced in the previous subsection will be useful for book keeping.

One additional transverse covariant derivative in the 2-gluon correlation function can be computed as follows. First, we again express gauge potentials in terms of field strengths
\begin{widetext}
\begin{eqnarray}
 \langle D^k A^i_{\underline a} (x^-,\vec x_\perp ) A^j_{\underline b}(y^-,\vec y_\perp)
  \rangle =
 \int_{-\infty}^{x^-} d {x'}^- \int_{-\infty}^{y^-} d {y'}^-  
  \left\langle (D^k  F^{+i})_{\underline a} (x'^-,\vec x_\perp ) F^{+j}_{\underline b} (y'^-,\vec
  y_\perp ) \right\rangle  \, .
\end{eqnarray}
\end{widetext}

Using the same change to covariant gauge as in Sec.\ \ref{sec:aa}, and recalling that 
$D_{\mathrm{cov}}^k F_{\mathrm{cov}}^{+i} = \partial^k \partial^i \alpha$, the expectation value on the right hand side can be transformed into the expression
\begin{widetext}
\begin{multline}
  \label{eq:daa}
  \left \langle \mathcal{U}^\dagger_{\underline a \underline a'} (x'^-, \vec x_\perp)
  \mathcal{U}^\dagger_{\underline b \underline b'} (y'^-, \vec y_\perp)
  \partial^k \partial^i \alpha_{\underline a'} (x'^-, \vec x_\perp) \partial^j
  \alpha_{\underline b'} (y'^-, \vec y_\perp) \right\rangle
  \\ = \delta_{\underline a \underline b} \frac{g^2}{N_c^2-1} \delta(x'^--  y'^-) \left[
  -\nabla_x^i \nabla_x^k \nabla_y^j \gamma(x'^-, \vec x_\perp, \vec y_\perp) \right]
  d(x'^-,  \vec x_\perp, \vec y_\perp) \, ,
\end{multline}
\end{widetext}
in analogy to Eq.\ (\ref{eq:FFgd}).  Note that correlators with three gluon fields vanish since an even number of adjoint Wilson lines and fields $\alpha$ have to be contracted with each other. Combinations $\langle U \alpha\rangle \sim 0$ are suppressed \cite{Fukushima:2007dy}. 

The two integrals over $x'^-$ and $y^-$ can be dealt with exactly as in the case of the simple 2-point function. The result for arbitrary longitudinal positions $x^- > 0$ (after taking the thickness of light cone sources to zero) is
\begin{multline}
  \langle D^k A^i_{\underline a} (\vec x_\perp ) A^j_{\underline b}(\vec x_\perp)
  \rangle \\
  = - \frac{g^2}{N_c^2-1} \delta_{\underline{ab}} \nabla_x^i \nabla_x^k \nabla_y^j \gamma(\vec
  x_\perp, \vec y_\perp) \Big|_{\vec y_\perp \to \vec x_\perp}  \,  ,
\end{multline}
in the interesting UV limit $r\to 0$. The same expectation value with the covariant derivative acting on the second gauge field would result in the same expression with the obvious replacement $\nabla_x^k \to \nabla_y^k$.

We apply the same basic strategy to calculate expressions with more derivatives. We obtain
\begin{multline}
  \label{eq:dada}
  \langle D^k A^i_{\underline a} (\vec x_\perp ) D^l
  A^j_{\underline b}(\vec x_\perp)
  \rangle \\
  = \frac{ g^2 \delta_{\underline{ab}}}{(N_c^2-1)}
  \nabla_x^i\nabla_x^k\nabla_y^j\nabla_y^l \gamma(\vec x_\perp, \vec y_\perp)
  \, .
\end{multline}
In the same spirit we have
\begin{multline}
\label{eq:ddaa}
  \langle D^k D^l A^i_{\underline a} (\vec x_\perp )
  A^j_{\underline b}(\vec x_\perp)
  \rangle \\
  = \frac{g^2 \delta_{\underline{ab}}}{(N_c^2-1)} \nabla_x^i\nabla_x^k\nabla_x^l\nabla_y^j \gamma(\vec x_\perp, \vec y_\perp)  \, .
\end{multline}

The higher derivatives of the correlation function $\gamma$ are straightforward to calculate. We have
\begin{multline}
  \nabla_x^i \nabla_x^k \nabla_y^j \gamma (\vec x_\perp, \vec y_\perp)
  =   \frac{\mu(\vec R)}{4\pi} \\ \times \left[ \left( \delta^{ij}\frac{r^k}{r} +
    \delta^{ik}\frac{r^j}{r} +  \delta^{jk}\frac{r^i}{r}  \right) mK_1(mr)  \right.
    \\  - \left.
    \frac{r^i r^j r^k}{r^3} m^2 r K_2(mr) \right]  \\ + \frac{\nabla^l \mu(\vec
    R)}{8\pi} \left( \delta^{jl} \frac{r^i r^k}{r^2} - \delta^{il}\frac{r^j
    r^k}{r^2} - \delta^{kl} \frac{r^i r^j}{r^2} \right) mr K_1(mr)
 \\  - \frac{\nabla^l \mu(\vec
    R)}{8\pi} \left( \delta^{jl} \delta^{ik} - \delta^{il}\delta^{jk} - \delta^{kl} \delta^{ij} \right) K_0(mr) \, ,
\end{multline}
where we kept the two leading orders, $1/r$ and $m$, in our power counting in $mr$. One can check that the contribution of the leading term to observables, such as $\beta^i$, vanishes due to the odd number of powers in $r^i$.  Hence the relevant term in the UV limit is
\begin{multline}
  \label{eq:dddg}
  \nabla_x^i \nabla_{x,y}^k \nabla_y^j \gamma (\vec x_\perp,
  \vec y_\perp)\big|_{\vec y_\perp \to \vec x_\perp}
  =   \frac{\nabla^l \mu(\vec R)}{16\pi}   \\ \times \ln \left( \frac{Q^2}{\hat m^2} \right)
       \left( \mp \delta^{jl} \delta^{ik} \pm \delta^{il}\delta^{jk} + \delta^{kl}
       \delta^{ij} \right)  \, .
\end{multline}
The lower signs are valid if the derivative $\nabla^k$ acts on $y_\perp$ instead of  $x_\perp$. The lower signs in the previous expression will be useful for the expectation value $\langle A^i_{\underline a} D^k A^j_{\underline b}(\vec x_\perp) \rangle $. As a consistency check, we note that Eq. (\ref{eq:dddg}) switches between upper and lower signs under the exchange $\{ i,\underline{a} \} \leftrightarrow \{ j,\underline{b} \}$ as dictated by symmetry.  As discussed above, we have dropped a term $\mathcal{O}(g^2\mu Q)$ that does not contribute to observables.

Caution is needed when calculating four derivatives acting on $\gamma$. The leading behavior of 
$\nabla_x^i \nabla_y^j \nabla_x^k \nabla_{x,y}^l \gamma (\vec x_\perp, \vec y_\perp)\big|_{\vec y_\perp \to \vec x_\perp}$ is similar to $\triangle \ln r$ which vanishes everywhere except for $r \rightarrow 0$. A proper integration will give us the leading term (again regularizing $1/r$ by $Q$) as
\begin{multline}
  \nabla_x^i \nabla_y^j \nabla_x^k \nabla_{x,y}^l \gamma (\vec x_\perp,
  \vec y_\perp)\big|_{\vec y_\perp \to \vec x_\perp}  \\  =
   \mp \frac{\mu(\vec R)}{32\pi} Q^2  \left(
    \delta^{ij}\delta^{kl} + \delta^{ik}\delta^{jl} + \delta^{jk}\delta^{il}
    \right) \, .
\end{multline}
In the UV limit the next to leading term in the transverse scale hierarchy is
\begin{multline}
  \nabla_x^i \nabla_y^j \nabla_x^k \nabla_{x,y}^l \gamma (\vec x_\perp,
  \vec y_\perp)\big|_{\vec y_\perp \to \vec x_\perp}   =
   \frac{\nabla^m \nabla^n \mu(\vec R)}{32 \pi} \\
    \times  \ln \left( \frac{Q^2}{\hat m^2} \right) (  \delta^{ij} \delta^{km} \delta^{ln} - \delta^{ik} \delta^{jm} \delta^{ln}
    \mp \delta^{il} \delta^{jm} \delta^{kn} \\ + \delta^{jk} \delta^{im} \delta^{ln} \pm \delta^{jl} \delta^{im} \delta^{kn}
    \mp \delta^{kl} \delta^{im} \delta^{jn} )  \\
    \pm  \frac{\nabla^m \nabla^n \mu(\vec R)}{96 \pi} \ln \left( \frac{Q^2}{\hat m^2} \right) \\
    \times  ( \delta^{ij} \delta^{km} \delta^{ln} + \delta^{ik} \delta^{jm} \delta^{ln}
    + \delta^{il} \delta^{jm} \delta^{kn} \\ + \delta^{jk} \delta^{im} \delta^{ln} + \delta^{jl} \delta^{im} \delta^{kn}
    + \delta^{kl} \delta^{im} \delta^{jn} )         \\
    \mp  \frac{\triangle \mu}{96 \pi } \ln \frac{Q^2}{\hat{m}^2}   (\delta^{ij} \delta^{kl} +\delta^{ik} \delta^{jl} +\delta^{il} \delta^{jk})      \, .
\end{multline}

The expectation values of gluon fields with up to two covariant derivatives will enable us to calculate the expectation values of components of the energy-momentum tensor up to order $\tau^2$ in the next section, including the effects of transverse flow. In addition, we will calculate energy density and pressure up to order $\tau^4$. To that end we also compute the leading $Q^4$ terms of the fourth order coefficients. However, we will neglect all effects of transverse gradients at fourth order, which would lead to very lengthy expressions. 

\section{The Energy-Momentum Tensor of Colliding Nuclei}
\label{sec:avemt}

After the discussion of gluon correlation functions in single nuclei we now return to the case of two colliding nuclei.  We will further break down the expressions for the components of the energy-momentum tensor in the small $\tau$ expansion in terms of the fields $A_1^i$ and $A_2^i$ in the individual nuclei. It is then straightforward to apply the results of the last section.

\subsection{Energy Density and Flow}

The expectation value of the initial energy density $\varepsilon_0$ from Eq. (\ref{eq:eps0}) can be written as \cite{Lappi:06}
\begin{multline}
  \varepsilon_0 \ \equiv \langle \varepsilon_0 \rangle
   = \frac{g^2}{2} f_{\underline{abe}}f_{\underline{cde}}   \\ \times
     \left(\delta^{ij}\delta^{kl}+\epsilon^{ij}\epsilon^{kl}\right)
     \langle A_{1,\underline{a}}^i A_{1,\underline{c}}^k \rangle_{\rho_1}
     \langle A_{2,\underline{b}}^j A_{2,\underline{d}}^l \rangle_{\rho_2}    \, .
\end{multline}
Note that in this chapter we calculate only averages of components of the energy-momentum tensor and will henceforth suppress the symbol $\langle \ldots \rangle$ in the notation for simplicity.  Applying (\ref{eq:AA3}) for each nucleus, the initial energy density is
\begin{equation}
 \varepsilon_0 (\vec x_\perp) = 
\frac{2 \pi N_c \alpha_s^3}{ N_c^2-1 } \mu_1 (\vec x_\perp) \mu_2 (\vec x_\perp)
\ln \left(\frac{Q_1^2}{\hat m^2}\right) \ln \left(\frac{Q_2^2}{\hat m^2}\right) \,,
\label{ini_den}
\end{equation}
where $\mu_1$ and $\mu_2$ are the expectation values of the densities of charges in nuclei 1 and 2, respectively, and $Q_1$ and $Q_2$ are UV scales chosen
for the wave function of nucleus 1 and 2, respectively. We have dropped terms proportional to $\nabla \nabla \mu /m^2$ which are subleading for the energy density. 

Expression (\ref{ini_den}) is very interesting.   The appearance of $\alpha_s$ to the power 3 can be understood in the  following way.  This classical calculation corresponds to the emission of a  gluon from source 1, the emission of another gluon from source 2, followed by  their fusion via a triple gluon vertex.  This involves 3 powers of the  coupling $g$ in the amplitude, hence to a power of 3 in $\alpha_s$ when the  amplitude is squared to get the energy density.  The initial energy density is very sensitive to the numerical value of $\alpha_s$, since changing it by a factor of 2 results in a change in the initial energy density of a factor of 8.  Quantum corrections to the classical CGC results are difficult to compute \cite{quantum_corrections2,Gelis:2008rw} but may change this sensitivity dramatically. For example, it is reasonable to expect that one  coupling is evaluated at the scale $Q_{1}$, the second coupling at the scale  $Q_{2}$, and the third at a common scale $Q$. Using the lowest order renormalization group result for the running coupling
\begin{equation}
\alpha_s(M^2) = \frac{1}{\beta_2 \ln (M^2/\Lambda^2_{\rm QCD})}
\end{equation}
with $\beta_2 = (11N_c - 2N_f)/12\pi$ we would get
\begin{equation}
 \varepsilon_0 (\vec x_\perp) \approx 
\frac{2\pi  N_c \alpha_s(Q^2)}{ \beta_2^2 (N_c^2-1) } \,  
\mu_1 (\vec x_\perp) \mu_2 (\vec x_\perp) \, .
\end{equation}
This triumvirate of running couplings is reminiscent of what happens when computing quantum corrections to the small-$x$ evolution of the gluon distribution
\cite{quantum_corrections1}. It appears that scale dependences are weaker once quantum corrections are established. Of course, the functions 
$\mu_i(\vec x_\perp)$ also depend to some degree on the scales.

For phenomenological purposes we will introduce a common UV scale $Q$ and, in what follows, we will always make the simplification 
$\log(Q_i^2/\hat m^2) \to \log(Q^2/\hat m^2)$, $i=1,2$.  For example, we can choose $Q^2$ to be simply the arithmetic mean of the scales of both nuclei, $Q^2=(Q_1^2+Q_2^2)/2$. This is a very good approximation in the traditional MV setup where nuclei are considered homogeneous slabs of color charges. For most realistic applications this will still be a reasonable choice. Recall that $Q_i$ is proportional to the saturation scale, $Q_i=K Q_{si}$ for a given nucleus $i$, with a numerical factor $K$. In collisions of two nuclei the relevant scale for the energy density is typically the larger of the two saturation scales 
\cite{Dumitru:2001ux,Lappi:2006xc}. However, experimentally accessible saturation scales do not cover a large range. Even for the largest nuclei at LHC energies they are at most a few GeV, barely one order of magnitude larger than $\Lambda_\mathrm{QCD}$. Hence, assuming one common scale from some averaging procedure between both nuclei seems sufficient for many purposes. Because of the limited range in $Q_s$ we will also neglect a dependence of $Q$ on the transverse coordinate which is, in principle, present. Thus we will not evaluate any transverse derivatives acting on $Q$.

The expectation value of the rapidity-even flow vector in the transverse direction at order $\tau$ is simply given by
\begin{eqnarray}
  \alpha^i  &=&  - 2\pi \alpha_s^3\frac{N_c}{ N_c^2-1 } \ln^2 \frac{Q^2}{\hat
    m^2} \nabla^i\left( \mu_1 \mu_2 \right) \\
   &=& -\varepsilon_0  \frac{\nabla^i\left( \mu_1 \mu_2 \right)}{\mu_1 \mu_2} \, .
 \end{eqnarray}
Separation of contributions from both nuclei for the rapidity-even flow vector leads to
\begin{eqnarray}
  \beta^i &=& g^2 f_{\underline{abe}} f_{\underline{cde}} \epsilon^{ij} \left( \epsilon^{mn}
  \delta^{kl} - \epsilon^{kl} \delta^{mn} \right) \\ 
&& \times  \left[ \langle (D^i A_{1,\underline{a}}^m) A_{1,\underline{c}}^k
   \rangle \langle A_{2,\underline{b}}^n A_{2,\underline{d}}^l \rangle
   \right. \\  
&& + \left. \langle   (A_{1,\underline{a}}^m) A_{1,\underline{c}}^k
  \rangle \langle (D^i A_{2,\underline{b}}^n) A_{2,\underline{d}}^l \rangle
  \right] \, .
\end{eqnarray}
The expectation value then takes a form complementary to $\alpha^i$
\cite{Chen:2013ksa}
\begin{equation}
  \beta^i = - \varepsilon_0 \frac{\mu_2 \nabla^i \mu_1 - \mu_1 \nabla^i
    \mu_2}{\mu_1\mu_2}   \, .
\end{equation}
Note that the expectation value of $\beta^i$ disappears for $\mu_1=\mu_2$.  Thus it vanishes for collisions of identical nuclei with impact parameter $b=0$. We have discussed in detail in \cite{Chen:2013ksa} how $\beta^i$ describes a rotation of the fireball for $b\ne 0$ while still preserving boost-invariance. We will come back to this in the next section.

\subsection{Higher Orders in $\mathbf{\tau}$}

The expectation values of terms at order $\tau^2$ can be calculated in a straight forward but increasingly lengthy manner.  For the coefficient $\delta$ we have the intermediate result
\begin{eqnarray}
    \delta  &=& \langle [D^m,E_0][D^m,E_0]+[D^m,B_0][D^m,B_0] \rangle \nonumber \\
     &=& g^2 f_{\underline{abe}}f_{\underline{cde}}
     \left(\delta^{ij}\delta^{kl}+\epsilon^{ij}\epsilon^{kl}\right) \nonumber \\
  &&   \times (\langle (D^m A^i)_{1,\underline{a}} (D^m A^k)_{1,\underline{c}} \rangle_{\rho_1}
     \langle A_{2,\underline{b}}^j A_{2,\underline{d}}^l \rangle_{\rho_2}  \nonumber \\
 &&    + \langle  A_{1,\underline{a}}^i  A_{1,\underline{c}}^k \rangle_{\rho_1}
     \langle (D^m A^j)_{2,\underline{b}} (D^m A^l)_{2,\underline{d}} \rangle_{\rho_2}  \nonumber \\
&&     + \langle (D^m A^i)_{1,\underline{a}} A^k_{1,\underline{c}} \rangle_{\rho_1}
     \langle A_{2,\underline{b}}^j (D^m A)_{2,\underline{d}}^l \rangle_{\rho_2} \nonumber \\
 &&    + \langle A^i_{1,\underline{a}} (D^m A^k)_{1,\underline{c}} \rangle_{\rho_1}
     \langle (D^m A)_{2,\underline{b}}^j A_{2,\underline{d}}^l \rangle_{\rho_2})  \nonumber \\
&&     + g^4 f_{\underline{abc}}f_{\underline{cde}} f_{\underline{fgh}}f_{\underline{hie}}
     \left(\delta^{ij}\delta^{kl}+\epsilon^{ij}\epsilon^{kl}\right) \nonumber  \\
&&     \times \big(\langle A_{1,\underline{a}}^m A_{1,\underline{f}}^m  A_{1,\underline{d}}^i A_{1,\underline{i}}^k \rangle_{\rho_1}
      \langle A_{2,\underline{b}}^j A_{2,\underline{g}}^l \rangle_{\rho_2} \nonumber \\
&&      +\langle A_{1,\underline{b}}^i A_{1,\underline{g}}^k \rangle_{\rho_1}
      \langle A_{2,\underline{d}}^j A_{2,\underline{i}}^l A_{2,\underline{a}}^m A_{2,\underline{f}}^m \rangle_{\rho_2} \big)    \, .
\end{eqnarray}
Using the higher twist gluon correlation function we derived in Sec. \ref{hightwist} this evaluates to
\begin{eqnarray}
\delta &=& 4 \varepsilon_0 Q^2 \ln^{-1} \left( \frac{Q^2}{\hat m^2} \right) \nonumber \\
&&+ \varepsilon_0 \left[1 + \frac{2Q^2}{3m^2} \ln^{-2} \left( \frac{Q^2}{\hat m^2} \right) \right] 
\left[ \frac{\triangle \mu_1} {\mu_1} + \frac{ \triangle \mu_2 }{\mu2} \right] 
 \nonumber  \\
&&+14\pi \alpha_s^2 \frac{N_c}{(N_c^2-1)} \varepsilon_0 \ln \left(\frac{Q^2}{\hat m^2}\right) (\mu_1 +  \mu_2)  \nonumber \\
&&+ \frac{14 \pi \alpha_s^2 }{3m^2} \frac{N_c}{(N_c^2-1)} \varepsilon_0 \Bigg[ 2 (\triangle \mu_1 + \triangle \mu_2)  \nonumber \\
&& + \frac{\mu_1^2 \triangle \mu_2 + \mu_2^2 \triangle \mu_1 }{\mu_1  \mu_2} \Bigg] \, .
\end{eqnarray}
The other coefficients at order $\tau^2$ are
\begin{eqnarray}
\omega &=& \frac{\varepsilon_0}{4 \mu_1 \mu_2}\big[ \nabla^1 \nabla^1 (\mu_1 \mu_2) -\nabla^2 \nabla^2 (\mu_1 \mu_2)
\nonumber \\
&& + 2 (\nabla^1 \mu_1 \nabla^1 \mu_2 - \nabla^2 \mu_1 \nabla^2 \mu_2)  \big]   \nonumber \\
&& -  \frac{ N_c} {N_c^2-1 } \frac{ 5 \pi \alpha_s^2 \varepsilon_0}{3 m^2}    \nonumber \\
&& \times \big[ - \nabla^1 \nabla^1 ( \mu_1 + \mu_2) + \nabla^2 \nabla^2 
( \mu_1 + \mu_2) \big]   \, , 
\label{eq:omega} \\
\gamma &=&\frac{\varepsilon_0}{2 \mu_1 \mu_2} \big[ \nabla^1 \nabla^2 (\mu_1 \mu_2) + (\nabla^1 \mu_1 \nabla^2 \mu_2 + \nabla^2 \mu_1 \nabla^1 \mu_2)  \big]   \nonumber \\
&& + \frac{ N_c} {N_c^2-1 } \frac{ 10 \pi \alpha_s^2 \varepsilon_0}{3 m^2}  \nabla^1 \nabla^2 ( \mu_1 + \mu_2)     \, .
\end{eqnarray}
It is interesting to note the hierarchy for terms at order of $\tau^2$.  Terms with derivatives are subleading to terms without, for example
$(\tau Q)^2 \gg (\tau Q)^2 (\nabla\nabla\mu)/(m^2 \mu) \gg \tau^2 (\nabla\nabla\mu)/\mu$, while true non-abelian terms of order $(\tau \alpha_s)^2 \mu$ could be large as well.

The energy flow $\xi^i$ at order $\tau^3$ can be expressed with the help of Eq.\ (\ref{eq:xi}) as derivatives of second order quantities.  The leading $Q^2$ correction to rapidity-odd flow at order $\tau^3$ is
\begin{align}
\zeta^i  = & - \frac{9}{2}\varepsilon_0 Q^2 \ln^{-1} \left(\frac{Q^2}{\hat m^2}\right)
 \frac{\mu_1 \nabla^i \mu_2 - \mu_2 \nabla^i \mu_1 }{\mu_1 \mu_2}  \nonumber \\
\end{align}
up to second order in transverse gradients.

At fourth order in $\tau$ we focus on the leading $Q^4$ contributions for simplicity. For the relevant coefficients we obtain the expectation values
\begin{align}
  \rho &= \frac{3}{32} \varepsilon_0 Q^4 \ln^{-1} \left( \frac{Q^2}{\hat m^2}\right) + 
\frac{3}{8}\varepsilon_0 Q^4 \ln^{-2} \left( \frac{Q^2}{\hat m^2}\right) \, ,  \\
  \kappa &=-\frac{64}{3}\rho  \, .
\end{align}

\section{Phenomenology of Classical Fields in Heavy Ion Collisions}
\label{sec:6}

With the results from the last section we are now ready to discuss the early time evolution of key quantities in high energy nuclear collisions analytically. We can compare to some numerical results available in the literature.

\subsection{Time Evolution of Energy Density and Pressure}
\label{sec:homogeneous}

Let us begin by first considering the very simple case of homogeneous, equally thick nuclei, in other words, the case of colliding slabs with $\mu_1=\mu_2$ being constants. In that case any dynamics comes solely from the longitudinal expansion of the system. Because of its simplicity, this is an approximation often employed in the literature to study the general behavior of color glass systems.

Neglecting transverse gradients, and keeping only the leading $(\tau Q)^k$ terms, the results from the last section imply
\begin{align}
\varepsilon &=T^{00}(\tau,\eta)=\varepsilon_0 \Big[ 1- \frac{(Q\tau)^2}{a} \left(1-\frac{1}{2}\cosh 2\eta \right) \nonumber \\
  &+  \frac{3(Q\tau)^4}{32a^2} (a+4) \left( 1-\frac{2}{3}\cosh 2\eta \right)  \nonumber \\ &+  \mathcal{O}(\tau^6) \Big] \, , \\
p_T&=T^{ii}(\tau,\eta)= \varepsilon_0 \Big[ 1- \frac{(Q\tau)^2}{a}  \nonumber \\
  &+  \frac{3(Q\tau)^4}{32a^2} (a+4) +  \mathcal{O}(\tau^6) \Big]   \, , \label{eq:pocket1} \\
p_L&=T^{33}(\tau,\eta)= -\varepsilon_0 \Big[ 1- \frac{(Q\tau)^2}{a} \left(1+\frac{1}{2}\cosh 2\eta \right) \nonumber \\
  &+  \frac{3(Q\tau)^4}{32a^2} (a+4) \left( 1+\frac{2}{3}\cosh 2\eta \right)  \nonumber \\ &+  \mathcal{O}(\tau^6) \Big]
 \, ,
\end{align}
where we have defined $a=\ln (Q^2/\hat m^2)$ for brevity.  We have neglected terms of order $(\alpha_s a)^2 Q^2 \mu$ for two reasons. First, we have not computed the corresponding terms for the 4th order in time so we cannot evaluate these terms consistently. The calculation is somewhat tedious and reserved for a future publication. Secondly, the exact relation between $\mu_i$ and $Q$ is not fixed from first principles. We can estimate that with a reasonable value $K=2$ the corrections to $p_T/\varepsilon$ and $p_L/\varepsilon$ are small up to about $Q\tau \approx 0.8$, which leads us to believe that the following analysis is valid.

We can write down very simple but powerful pocket formulas for the time evolution of key quantities. For example the transverse and longitudinal pressure relative to the energy density at midrapidity behave as
\begin{align}
  \frac{p_T}{\varepsilon} (\tau)=& \frac{1-\frac{1}{a}(Q\tau)^2 + \frac{3(a+4)}{32a^2}(Q\tau)^4}{1-\frac{1}{2a}(Q\tau)^2 + \frac{a+4}{32a^2}(Q\tau)^4} \, , \nonumber \\
  \frac{p_L}{\varepsilon} (\tau)=& -\frac{1-\frac{3}{2a}(Q\tau)^2 +
    \frac{5(a+4)}{32a^2}(Q\tau)^4}{1-\frac{1}{2a}(Q\tau)^2+ \frac{a+4}{32a^2}(Q\tau)^4} \, .
  \label{eq:pocket}
\end{align}
Suppose we drop the order $\tau^4$ terms in the numerator and denominator.  Then at a time given by 
$(Q\tau)^2 = \frac{4}{5} a$ we have $p_T = p_L = \frac{1}{3} \varepsilon$.  This corresponds to the equation of state of a massless gas of quarks and gluons. 

\begin{figure}[b]
  \includegraphics[width=0.99\columnwidth]{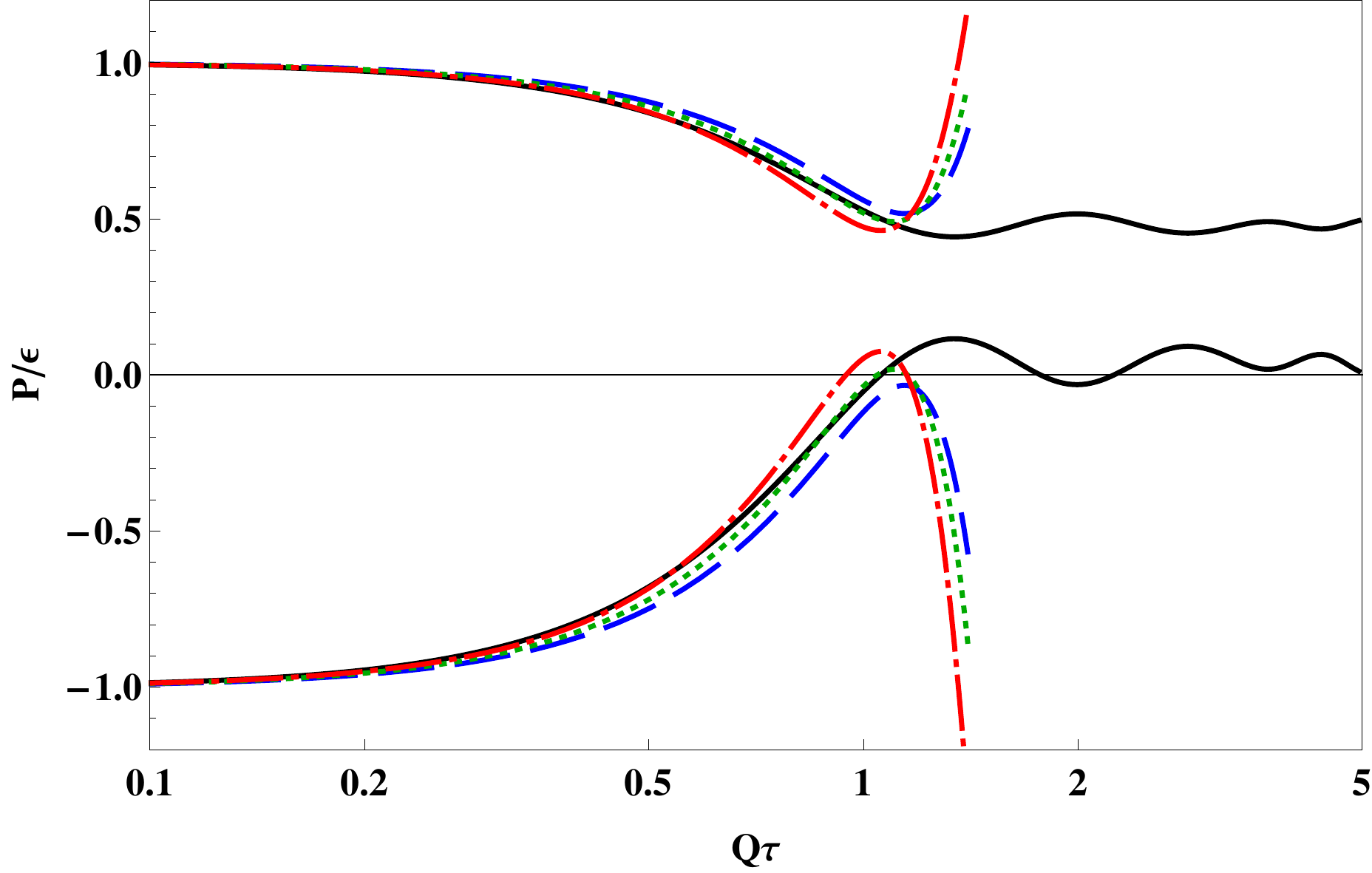}
  \caption{(color online) Evolution of the ratios of the transverse (upper
    curves) and longitudinal (lower curves) pressure over energy density for
    the classical gluon field at fourth order accuracy in time, compared to
    the numerical result from \cite{Gelis:2013rba} at leading order 
    for $g=0.5$ (solid
    lines). Values of $a$ = 0.8, 0.9, and 1.0 (dash-dotted, dotted and dashed
    lines resp.) are indicated.}
  \label{fig:numerics}
\end{figure}

\begin{figure*}[t]
\begin{center}
\includegraphics[width=0.82\linewidth]{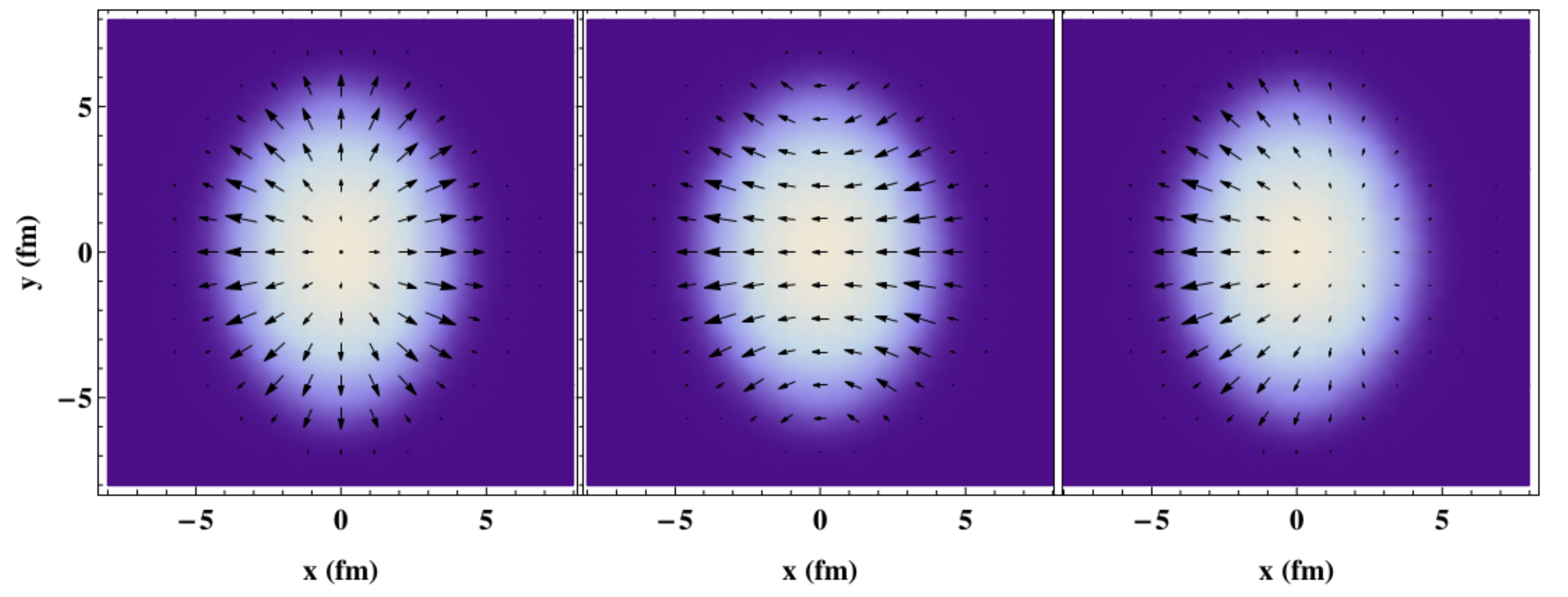}
  \caption{(color online) Different flow fields (black arrows) and initial energy density $\varepsilon_0$
     (shading) for Pb+Pb collisions at impact parameter $b=6$ fm in the 
     $x-y$-plane. The nucleus centered at $x=3$ fm travels in the positive 
     $\eta$-direction. Left panel: $\alpha^i$. Center Panel: $\beta^i$.
     Right Panel: Full transverse Poynting vector $T^{0i}$ at $\eta=1$. Note
     that $\alpha^i$ is proportional to $T^{0i}$ at $\eta=0$.}
\end{center}
  \label{fig:PbPbflow}
\end{figure*}

We can compare the results in Eq. (\ref{eq:pocket}) with those of Gelis and Epelbaum \cite{Gelis:2013rba}.  They performed a real-time lattice simulation
for colliding slabs using the gauge group $SU(2)$.  In Fig. \ref{fig:numerics} we show results for the transverse and longitudinal pressures over the energy
density, $p_T/\varepsilon$ and $p_L/\varepsilon$, from our analytic approach up to fourth order in $\tau$ and the numerical results from \cite{Gelis:2013rba}
(labeled LO in their work). Here we have chosen the values of $a$ = 0.8, 0.9, and 1.0, all of which give very good matching for small time, and are not unreasonable for small saturation scales. Note that this is a very schematic comparison for several reasons. A more quantitative statement would require a careful analysis of the IR and UV scales in the numerical calculation and their relation to $Q_s$, a further investigation of $\mu^3$ and $\mu^5$ terms in the analytic result, and the use of $SU(2)$ instead of $SU(3)$ in our calculations. However, it is interesting to note that the results agree quite well up to $\tau \sim 1/Q$.

The comparison with numerical work is important in two ways. First, the study in \cite{Gelis:2013rba} indicates that classical field dynamics is sufficient for times smaller than $\tau_0 \sim 1/Q_s$, at least at small to moderate values of the strong coupling $g$. After that time quantum corrections and instabilities start to dominate. The successful comparison also validates our previous argument about the convergence radius of the small-time expansion which we expected to be given parametrically by $1/Q_s$. Indeed we can reproduce the results for transverse and longitudinal pressure very well up to that time.  If we would want to relax the conditions and allow transverse gradients, we would also introduce dimensionless terms $\tau \nabla^i$ which are smaller than $\tau Q$ in the region of applicability.

Serendipitously our near-field expansion works rather well up to the same time scale to which the classical field approach is valid. Thus we are led to believe that our analytic results are a rather simple and almost complete account of the collision dynamics up to $\tau_0$.  The asymptotic values for $p_T/\varepsilon$ and $p_L/\varepsilon$ reached in the classical theory after $\tau_0$ are $\sim 1/2$ and $\sim 0$, respectively.  Quantum corrections and instabilities will, however, lead to further isotropization soon after $\tau_0$ \cite{Gelis:2013rba}.

\subsection{Global Flow of Glasma}
\label{sec:flow}

Two of us have discussed the effect of the two first-order flow terms $\alpha^i$ and $\beta^i$ in detail in \cite{Chen:2013ksa}.  The hydrodynamic-like flow term $\alpha^i$ obviously leads to both radial and elliptic flow; see left panel of Fig. 6. Note that this is flow of energy of the classical gluon field at this point. However, due to energy and momentum conservation, this flow will translate into a flow of fluid cells after thermalization.  We will discuss this in a future publication.  

\begin{figure}[h*]
\begin{center}
\includegraphics[width= 0.8\linewidth]{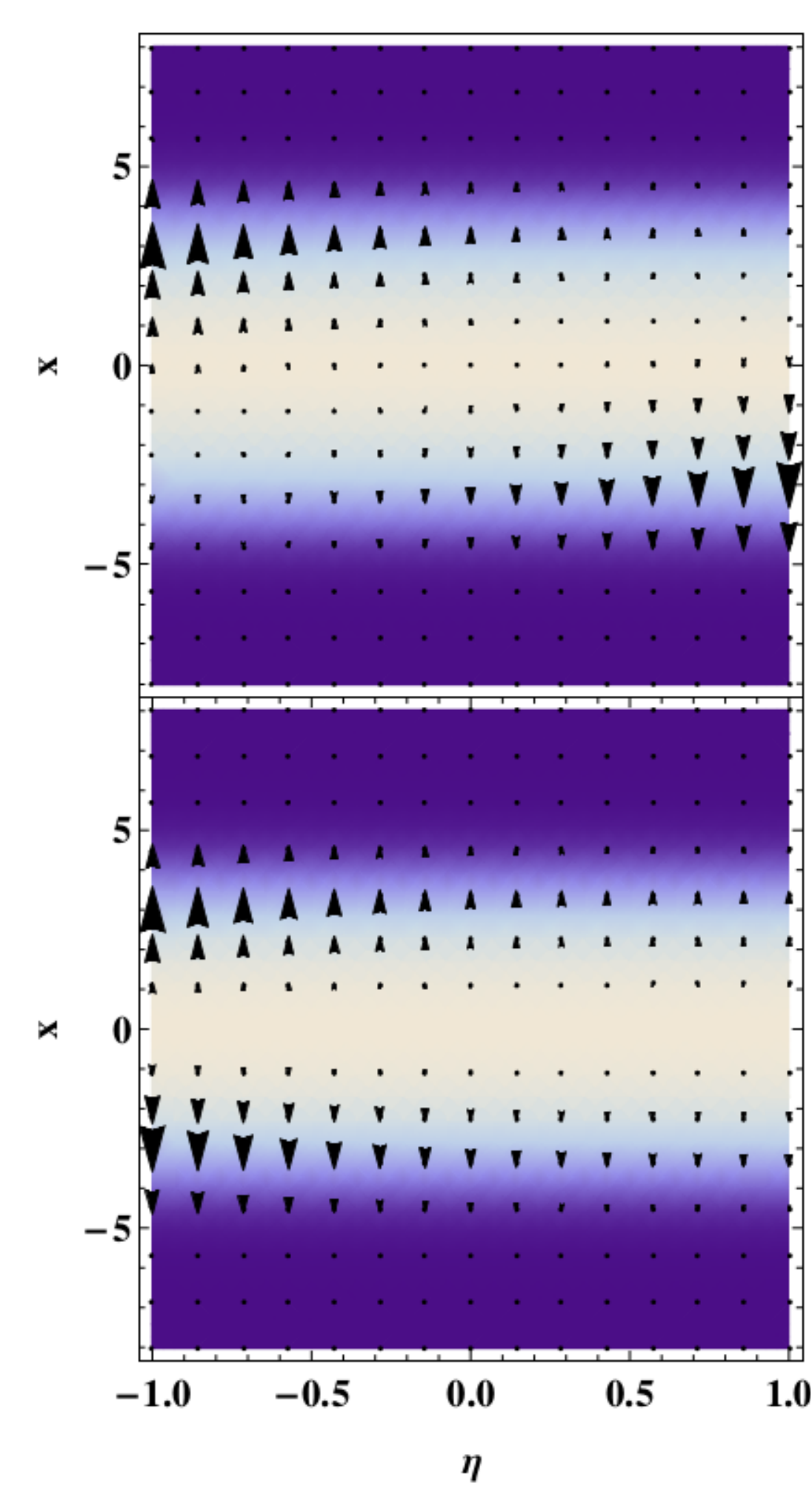}
\end{center}
\caption{(color online) Transverse Poynting vector $T^{0i}$ (black arrows) and initial energy density $\varepsilon_0$ (shading) in the $\eta-x$ plane at $y=0$.  Top panel: Pb+Pb at $b=6$ fm. Angular momentum is carried by the gluon  field. Bottom panel: Pb+Ca at $b=0$ fm, the Pb nucleus is moving to the right. The system expands more strongly in the wake of the larger nucleus.}
 \label{PbCaflow}
\end{figure}

The rapidity-odd flow term $\beta^i$ potentially has many interesting implications; see center panel of Fig. 6. Its event average vanishes for central collisions (impact parameter $b=0$) for collisions of identical nuclei. However for finite impact parameters it carries the angular momentum of the gluon field that is transferred from the non-vanishing angular momentum of the two colliding nuclei. The flow field exhibits a characteristic rotation pattern around the impact vector; see the right panel of Fig. 6 and top panel of Fig. \ref{PbCaflow}. This would lead to directed flow $v_1$ of particles which has been observed in experiments. The angular momentum would be transferred to the quark-gluon fluid at a later stage with potential interesting consequences \cite{Liang:2004ph,Csernai:2013bqa}.  We again refer the reader to \cite{Chen:2013ksa} for more details. In collisions of two different species of nuclei, $\beta^i$ leads to an increase of the radial flow in the wake of the larger nucleus while suppressing flow in the wake of the smaller nucleus; see bottom panel of Fig. \ref{PbCaflow}. For asymmetric collisions at finite impact parameter the flow field becomes more complicated.  This could lead to interesting flow patterns unique to classical gluon field dynamics \cite{Chen:2013ksa}. Those could be a novel signature for the importance of color glass condensate in this regime.  For the illustrations shown here, Woods-Saxon profiles have been used for the volume density of nucleons in the nuclei from which the transverse color charge densities $\mu_{1,2}$ are computed.

The second order in time also introduces a pressure anisotropy in the transverse plane for asymmetric collision systems. The eccentricity of the transverse pressure $\epsilon_p = (T^{11}-T^{22})/(T^{11}+T^{22})$ is often used to measure the buildup of elliptic flow in the system.  For the event average we read off from 
Eq. (\ref{eq:tii}) that 
\begin{equation}
  \epsilon_p(x,y,\eta) = \frac{\omega(x,y)\tau^2}{4\varepsilon_0(x,y)} \, ,
\end{equation}
up to second order in gradients and up to second order in $\tau$.  This quantity is independent of $\eta$.  We see that the pressure anisotropy indeed starts to grow quadratically in time.  We leave further numerical analysis to a future paper.

\begin{figure}[tb]
  \begin{center}
  \includegraphics[width=1.0\columnwidth]{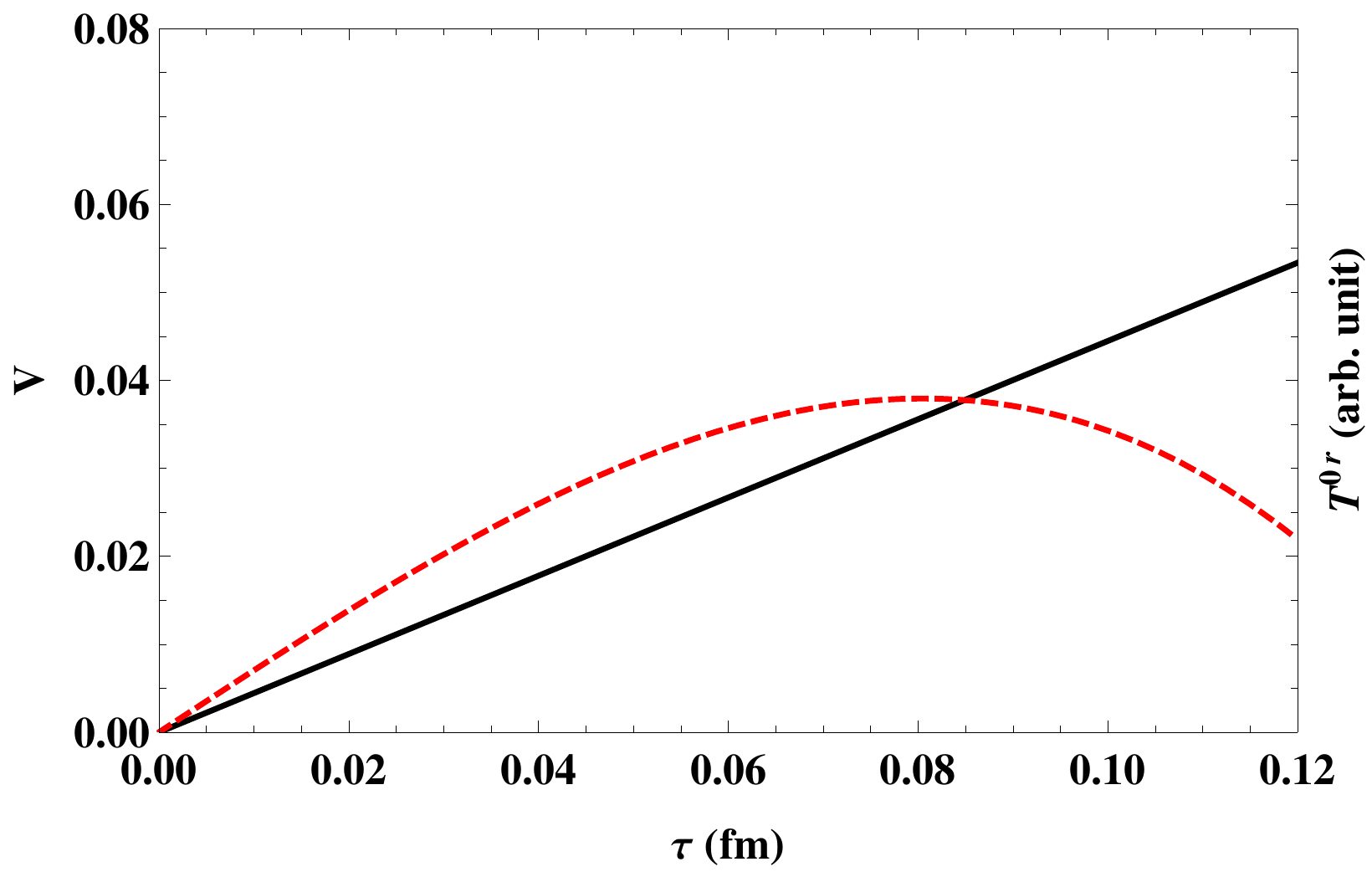}
  \end{center}
  \caption{(color online) Time evolution for $|V|$ (solid black line) and for the radial 
  projection of $T^{0i}$ (arbitrary units, dashed red line) for central Pb+Pb
  collisions at midrapidity with approximations as described in the text.}
  \label{fig:9}
\end{figure}

\begin{figure}[tb]
  \begin{center}
  \includegraphics[width=0.91\columnwidth]{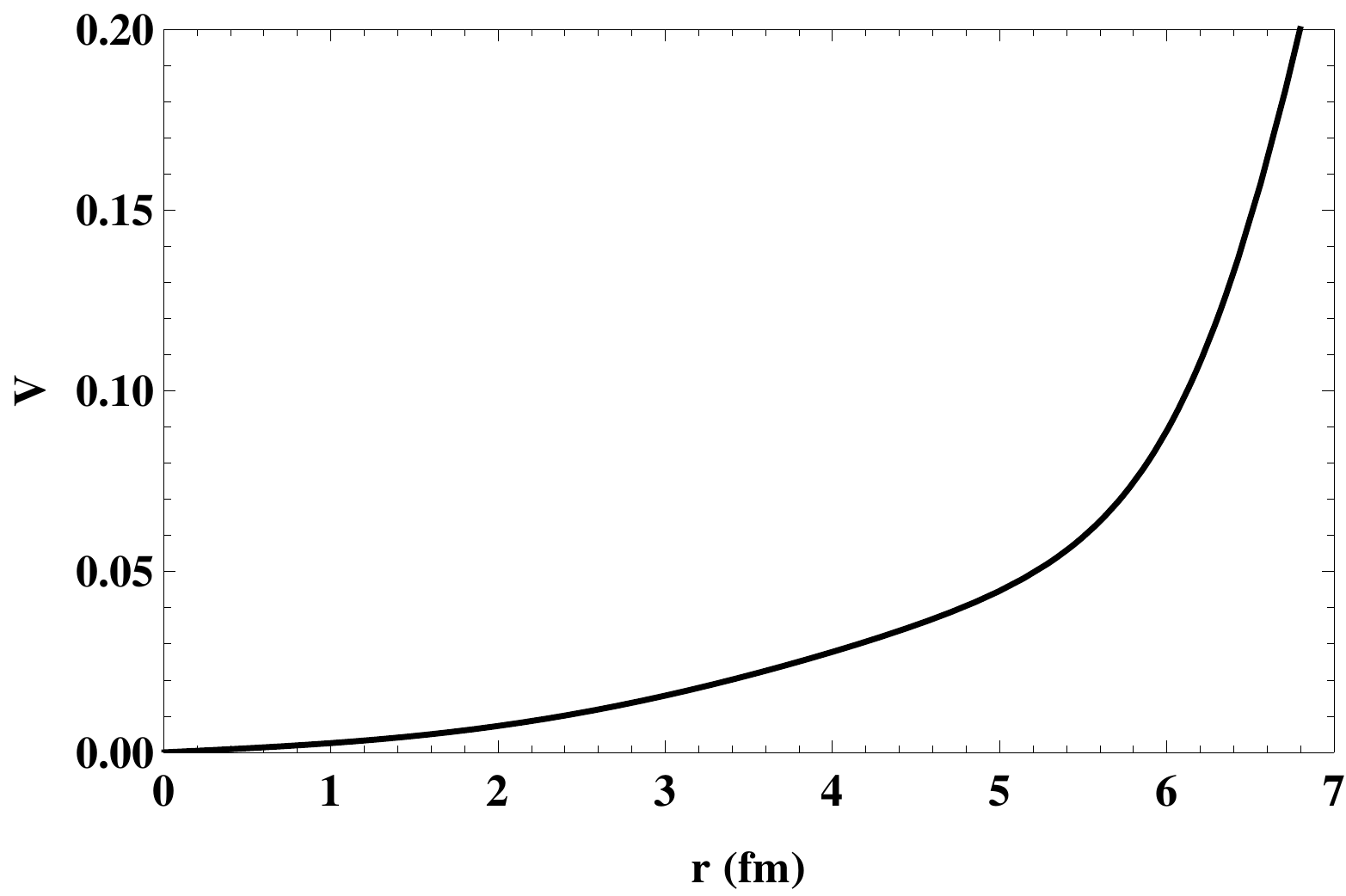}
  \end{center}
  \caption{Dependence of $|V|$ on the radial coordinate $r$ for central Pb+Pb
  collisions at midrapidity. The radius of a Pb nucleus is about 7 fm.}
  \label{fig:10}
\end{figure}

Third order corrections typically slow the linear growth of the energy flow.  For the rapidity-even part we again have a compact formula if we neglect
terms with three or more derivatives. From Eq. (\ref{eq:xi}) and the expression for the expectation values of $\alpha^i$ and $\delta$ we obtain
\begin{equation}
  T^{0i}_{\mathrm{even}} =
  \frac{\tau}{2} \, \alpha^i \left( 1 -\frac{1}{2a} 
  (Q\tau)^2  \right)  \cosh\eta \, .
\end{equation}
Similarly, from the expectation values for $\beta^i$ and $\zeta^i$, we have
\begin{equation}
  T^{0i}_{\mathrm{odd}} =
  \frac{\tau}{2} \, \beta^i \left( 1- \frac{9}{16a} 
  (Q\tau)^2  \right) \sinh\eta \, ,
\end{equation}
when higher order gradients and terms or order $\mu_i^3$ are neglected.  Interestingly, when we look at $V^i = T^{0i}/T^{00}$ at midrapidity as a proxy for velocity, the leading corrections in the time evolution cancel in numerator and denominator. They are of order $-(Q\tau)^2/2a$ for both the energy density and the even part of the energy flow. In other words, while the growth of $T^{0i}$ slows and invariably peaks and diminishes due to the longitudinal expansion, the velocity $V^i$ continues to grow roughly linearly as
\begin{equation}
  V^i = -\frac{\tau}{2} \frac{\nabla^i \varepsilon_0}{\varepsilon_0} 
   + \mathcal{O}(\tau^5) \, ,
\end{equation}
at midrapidity when transverse gradients of third order and higher are neglected.  Figure \ref{fig:9} shows the time evolution of the radial velocity $|V|$ 
up to corrections of order $\tau^5$ for a point $r=5$ fm away from the center of a central Pb+Pb collision at midrapidity. We also computed the time evolution of the radial projection of $T^{0i}$, including the $\tau^3$ correction, to contrast its slowing down to the linear growth of $V^i$. For the calculation of $T^{0i}$, we have chosen $Q^2 = 2$ GeV$^2$ and $a=1$. Figure \ref{fig:10} displays the radial dependence of $|V|$ for the same central Pb+Pb collisions at $\tau=0.2$ fm. We see that the surface velocity peaks around 0.2. However, one has to be cautioned that typically the first fermi of the boundary (beyond $r=6$ fm for a Pb nucleus) is usually outside of the applicability of this calculation.

\subsection{Towards Quark Gluon Plasma}
\label{sec:spacetime}

Let us summarize our knowledge of nuclear collisions at a typical time $\tau_0 = 1/Q_s$. The energy-momentum tensor can be written, up to third order in $\tau$ as  
\begin{widetext}
\begin{equation}
  T^{m n} =
  \begin{pmatrix}
    \varepsilon_0-\frac{\tau^2}{8}(-2\triangle \epsilon_0 + \delta)  & \frac{\tau}{2} \alpha^x + \frac{\tau^3}{16} \xi^x & \frac{\tau}{2} \alpha^y + \frac{\tau^3}{16} \xi^y  & \frac{\tau}{8} \nabla^i \beta^i \\
    \frac{\tau}{2} \alpha^x + \frac{\tau^3}{16} \xi^x  & \varepsilon_0-\frac{\tau^2}{4}(-\triangle \varepsilon_0 + \delta-\omega) & \gamma & \frac{1}{2} \beta^x  \\
    \frac{\tau}{2} \alpha^y + \frac{\tau^3}{16} \xi^y  & \gamma & \varepsilon_0-\frac{\tau^2}{4}(-\triangle \varepsilon_0 + \delta+\omega) & \frac{1}{2} \beta^y   \\
    \frac{\tau}{8} \nabla^i \beta^i    & \frac{1}{2} \beta^x  & \frac{1}{2} \beta^y  & -\frac{\varepsilon_0}{\tau^2} + \frac{1}{8}(-2\triangle \varepsilon_0 + 3 \delta)
      \end{pmatrix} \, .
      \label{tmn}
\end{equation}
\end{widetext}
Here we have used the $\tau,x,y,\eta$ coordinate system for the tensor. This gets rid of unwieldy $\cosh \eta$ and $\sinh\eta$ terms from boosts. Note that there is no explicit dependence on $\eta$ in this coordinate system due to boost invariance.  This tensor exhibits the standard features expected of a fireball: radial and elliptic flow, and a decrease of energy density and pressure with time, mostly due to the longitudinal expansion. In addition, we find angular momentum and directed flow for finite impact parameter collisions, and a complicated flow pattern for asymmetric collision systems. These features can be predicted more or less accurately and in analytic form averaged over events.

The reader should keep in mind that the phenomenological analyses in the present section are rather crude and could be refined in many ways, as pointed out numerous times. However, they result in compact pocket formulas which could be useful for quick estimates in many situations. A more careful analysis can be done starting with the full expressions from Sec. \ref{sec:avemt}.

After a proper time $\tau_0$, instabilities growing from small fluctuations take over, leading to turbulent behavior of the fields. Further isotropization and equilibration is then expected to lead to quark-gluon plasma near kinetic equilibrium.  From a phenomenological perspective, one could simply translate the energy-momentum tensor of the classical field around the time $\tau_0$ directly into hydrodynamic fields, as was done in \cite{Fries:2005yc} for ideal hydrodynamics and in \cite{Gale:2012rq} for viscous hydrodynamics. However, this obviously leads to large shear stress corrections, as can be seen from the large difference between transverse and longitudinal pressure around $\tau_0$ as presented previously.  It would be very interesting to see how key features of the transverse flow field translate into hydrodynamics and how they fare during subsequent hydrodynamic evolution. This would enable us to connect features of classical gluon fields in the initial state to observables.

It would be relatively straight forward to build a semi-analytic event generator from our results.  For example, one could follow reference \cite{Schenke:2012wb} which used a model for charge configurations of nuclei in collisions. In our approach, their numerical solution to the Yang-Mills equations would be replaced by our analytic time evolution using the near-field approximation. Then, from the sampled charge distributions, one has to calculate the coefficients $\epsilon_0$, $\alpha^i$, $\beta^i$, $\delta$, etc.\ to obtain an event-by-event energy-momentum tensor.

\section{Conclusion}

In this paper we worked out analytic solutions of the Yang-Mills equations for two nuclei with random color charges colliding on the light cone.  Using a recursive solution we computed the early time gluon field and energy-momentum tensor in a near-field approximation. We find that this approximation gives acceptable results roughly up to a time $\tau_0$ given by the inverse of the saturation scale $Q_s$. This coincides with the time at which the entire classical field approximation starts to breaks down anyway. Explicit expressions for the fields and energy-momentum tensor up to order $\tau^4$ have been provided.

We have also calculated expectation values for the energy-momentum tensor when many events are averaged. Our calculation generalizes the McLerran-Venugopalan model to allow small but non-vanishing gradients in the average color charge in the transverse plane. This permitted us to discuss flow phenomena in averaged events.  We provide a comprehensive set of expectation values of coefficients of the energy-momentum tensor which allow predictions for event-averaged $T^{\mu\nu}$ for times around $\tau_0$.  We give compact and analytic formulas for key quantities like the time evolution of energy density, transverse and longitudinal pressure, the time evolution of transverse flow of energy, and the time evolution of the transverse pressure asymmetry.

We find that the transverse flow of energy grows linearly with time and that it can reach sizeable values at the surface of the fireball at $\tau_0$. We
have also discovered that the asymmetry between transverse pressures starts to grow quadratically in time. The time evolution of transverse and longitudinal pressure matches well with numerical results available in the literature up to $\tau_0$.  Besides the usual radial and elliptic flow a rapidity-odd flow emerges.  We suggest that this energy flow of the glasma could be the origin of directed flow. It carries angular momentum which rotates the fireball.  More complex flow patterns appear for collisions of asymmetric nuclei.  The characteristic glasma flow pattern could potentially lead to another signature for color glass dynamics in high energy collisions.

At $\tau_0 \sim 1/Q_s$ our calculation becomes unreliable. However, it could be attempted to match our results to a (3+1)-D viscous hydrodynamic code. We will discuss this in a forthcoming publication.  We have also discussed the possibility to construct an event generator based on the results of this paper.

\section*{Acknowledgement}
RJF and GC thank L. McLerran for discussion and encouragement, and RJF and JIK thank L. Csernai for comments on the manuscript.  We are grateful to M. Li for
checking many equations in the manuscript for errors and typos.  RJF and GC were supported by the U.S. National Science Foundation through CAREER grant
PHY-0847538, and by the JET Collaboration and Department of Energy grant DE-FG02-10ER41682. GC also acknowledges partial support from the US Department 
of Energy Grant No. DE-FG02-87ER40371. JIK and YL were supported by the Department of Energy grant  DE-FG02-87ER40328.

\begin{appendix}

\section{General Definitions}
\label{sec:app1}

Some conventions and useful formulae are gathered in this appendix.  3-vectors are denoted by bold symbols, vector arrows denote 2-vectors in the transverse plane.  As an example, $x^\mu = (t,\mathbf{x}) = (t,\vec x_\perp,z)$.  Light cone coordinates are defined by
\begin{equation}
  x^\pm = \frac{1}{\sqrt{2}} \left( x^0 \pm x^3\right)\, ,
\end{equation}
with $d^4 x = dx^+ dx^- d^2 x_\perp$ and $x^\mu y_\mu = x^+ y^- + x^- y^+ -
x_\perp^i y_\perp^i$. 
Note that $\partial^\mu = (\partial/\partial t, -\nabla)$ and
$\partial^\pm = \partial/\partial x^\mp$.
Unless indicated otherwise, small Latin indices $i,j,k$ indicate transverse components of a vector, Greek indices label 4-vectors in $(t,x,y,z)$ coordinates, and Latin indices $m,n$ label 4-vectors in $(\tau,x,y,\eta)$ coordinates.  Underlined Latin indices refer to the $SU(3)$ algebra.

Proper time $\tau$ and space-time rapidity $\eta$ for a space-time point $x^\mu$ are defined as
\begin{align}
  \tau & = \sqrt{t^2 - z^2} = \sqrt{2x^+x^-} \, , \\
  \eta & = \frac{1}{2} \ln \left( \frac{t +z}{t -z} \right) =
  \frac{1}{2} \ln \left( \frac{x^+}{x^-} \right) \, .
\end{align}
It is useful to express Cartesian and light cone derivatives via hyperbolic ones by
\begin{equation}
  \partial^\pm = \frac{x^\pm}{\tau} \frac{\partial}{\partial \tau} \mp
  \frac{1}{2x^\mp} \frac{\partial}{\partial \eta} \, ,
  \label{eq:der1}
\end{equation}
and
\begin{align}
  \frac{\partial}{\partial t} =& \cosh\eta \frac{\partial}{\partial \tau}
  - \frac{1}{\tau} \sinh\eta \frac{\partial}{\partial \eta} \, ,
  \label{eq:der2} \\
  \frac{\partial}{\partial z} =& -\sinh\eta \frac{\partial}{\partial \tau}
  + \frac{1}{\tau} \cosh\eta \frac{\partial}{\partial \eta}\, .
  \label{eq:der3}
\end{align}

Our conventions for covariant derivatives and field strength tensors are
\begin{align}
  D^\mu &= \partial^\mu - ig A^\mu  \, ,  \\
  F^{\mu\nu} &= \frac{i}{g} [D^\mu,D^\nu] = \partial^\mu A^\nu - \partial^\nu A^\mu -ig [A^\mu,A^\nu] \, .
\end{align}
Here $A_\mu$, $F_{\mu\nu}$ and $J_\mu$ are $SU(N_c)$ valued functions that can be expressed as linear combinations of the $SU(N_c)$ generators $t^{\underline{a}}$, $\underline{a}=1,\ldots,N_c^2-1$. The generators are defined through $[t^{\underline{a}},t^{\underline{b}}] = i f^{\underline{abc}} \,
t^{\underline{c}}$ and normalized by
\begin{equation}
  \tr t^{\underline{a}} = 0 \, , \quad \tr(t^{\underline{a}} t^{\underline{b}})
  = \frac{1}{2} \delta^{\underline{a}\underline{b}} \, .
\end{equation}
This immediately implies that
\begin{equation}
  \tr(X) = 0, \quad \tr([X,Y]) = 0 \, ,
\end{equation}
for any $X$, $Y$ in the $SU(N_c)$ algebra since $[X,Y] \in SU(N_c)$.

Using the ordinary product rule and the Jacobi identity, one can show that covariant derivatives obey the generalized product rule
\begin{equation}
  [D^i, XY] = [D^i X] Y + X[D^i, Y] \, ,
\end{equation}
for any $X$, $Y$ in the $SU(N_c)$ algebra; in particular
\begin{equation}
  [D^i, [X,Y]] = [[D^i X], Y] + [X,[D^i, Y]] \, .
\end{equation}

It is sometimes helpful to interpret $\tr(AB)$ as a bilinear scalar product on $SU(3)$ and $[A,B]$ as a skew-symmetric product whose result is orthogonal to both $X$ and $Y$ such that
\begin{equation}
  \tr(X,[X,Y]) = 0 \, .
\end{equation}
This leads to some important ways to simplify expressions with a trace involved.  They include:
\begin{eqnarray}
  \label{eq:apptrcov}
  \tr(X[D^i,X]) &=& - \frac{1}{2} \nabla^i \tr(X^2) \, ,\\
  \epsilon^{ij} [D^i,[D^j,X]] &=& ig [B^3,X] \label{eq:id2} \, ,\\
  \tr\left( [D^i X][D^i X] \right) &=& \frac{1}{2} \triangle \tr (X^2) \nonumber \\ 
&-& \tr\left( X [D^i,[D^i,X]] \right) \label{eq:id3} \, ,\\
  \tr\left( \epsilon^{ij} [D^i,X][D^j,Y]\right) &=&
  \frac{1}{2} \epsilon^{ij} \nabla^i \nonumber \\
 &\times& \tr\left([D^j,X]Y - [D^j,Y]X \right) \nonumber \\
     &+&\frac{1}{2} \epsilon^{ij} \tr ([D^i,[D^j,X]Y \nonumber \\
     &-&[D^i,[D^j,Y]X) \, ,\\
  \tr \left( X[D^i,[D^i,X]] \right) &=& \frac{1}{2} \triangle \tr (X^2) \nonumber \\ 
&-& \tr ([D^i,X][D^i,X]) \, , \label{eq:id6}
\end{eqnarray}
where $X$ and $Y$ are any $SU(N_c)$ fields.  As an example, the second equation implies that
\begin{align}
  \epsilon^{ij} \tr (E_0 [D^i,[D^j,B_0]] =& 0 \, , \label{theta1} \\
  \epsilon^{ij} \tr (B_0 [D^i,[D^j,E_0]] =& 0 \label{theta2} \, .
\end{align}

\section{Expressions at Order $\tau^3$ and $\tau^4$}
\label{sec:app2}

At order $\tau^3$ the transverse fields are
\begin{eqnarray}
  F^{i\pm}_{(3)} &=& -\frac{e^{\pm\eta}}{4\sqrt{2}}
  \left( [ D^j, F^{ji}_{(2)}] \pm [ D^{i}, F^{+-}_{(2)}] \right) \nonumber \\
  &+& \frac{ig}{8} \left( \epsilon^{ij} [B_0,F^{j\pm}_{(1)}] \pm
  [E_0,F^{i\pm}_{(1)}] \right) \nonumber \\
  &\mp& \frac{ig}{8}\frac{e^{\pm\eta}}{2\sqrt{2}} \epsilon^{ij}[ D^j,
  [E_0,B_0 ] ] \, ,
\end{eqnarray}
whereas $E^3_{(3)} = 0 = B^3_{(3)}$.  In terms of the initial fields the third order fields are
\begin{align}
  E^i_{(3)} =&  -\frac{1}{16} \left( \cosh\eta \, \epsilon^{ij}
  [D^j,[D^k,[D^k,B_0]]] \right. \nonumber \\
  & + \left. \sinh\eta \, [D^i, [D^k,[D^k,E_0]]] \right) \nonumber \\
  & - \frac{ig}{16}   \epsilon^{ij} \sinh\eta
  \left( [B_0,[D^j,E_0]] + [E_0, [D^j,B_0]] \right)  \nonumber   \\
  &- \frac{ig}{16}\cosh\eta  \left( [E_0,D^i,E_0 ] - [B_0,[D^i, B_0]]\right)
     \nonumber\\
  &-\frac{ig}{16} \sinh\eta \epsilon^{ij} [D^j, [E_0,B_0]]
  \, , 
\end{align}
and
\begin{align}
  B^i_{(3)} =& - \frac{1}{16} \left( \sinh\eta [D^i,[D^k,[D^k,B_0]]]
  \right. \nonumber \\  & \left.
  - \cosh\eta \epsilon^{ij} [D^j,[D^k,[D^k,E_0]]] \right) \nonumber \\
   &- \frac{ig}{16} \cosh\eta \left( [B_0,[D^i,E_0]] + [E_0,[D^i,B_0]]
   \right) \nonumber \\
   &- \frac{ig}{16}\sinh\eta \epsilon^{ij} \left( [B_0,[D^j,B_0]]
    - [E_0,[D^j,E_0]] \right) \nonumber \\
  &-\frac{ig}{16} \cosh\eta [D^i, [E_0,B_0]] \, .
\end{align}

The longitudinal field at order  $\tau^4$ is
\begin{align}
E^3_{(4)}=&\frac{1}{64}[D^i,[D^i,[D^j,[D^j,E_0]]]]   \nonumber \\
& +\frac{1}{16}ig \epsilon^{ij}[[D^i,E_0],[D^j,B_0]] \, ,
\end{align}
\begin{align}
B^3_{(4)}=&\frac{1}{64}[D^i,[D^i,[D^j,[D^j,B_0]]]]  \nonumber \\
 &
-\frac{1}{64}ig \epsilon^{ij}[[D^i,E_0],[D^j,E_0]] \nonumber \\
&+ \frac{3}{64}ig \epsilon^{ij}[[D^i,B_0],[D^j,B_0]] \nonumber \\
&+ \frac{g^2}{64} [E_0,[B_0,E_0]]   \, .
\end{align}

For the energy-momentum tensor the transverse flow vectors $\xi^i$ and $\zeta^i$, as defined in Eq. (\ref{eq:xizeta}), are given in terms of $E_0$ and $B_0$ by
\begin{eqnarray}
  \xi^i &=&  \left[D^i ,E_0 [ D^l,[ D^l, E_0]]
    + B_0 [D^l,[D^l,B_0]] \right] \nonumber \\
    &+& [D^i,E_0][D^l,[D^l,E_0]] + [D^i,B_0] [D^l,[D^l,B_0]] \nonumber \\
    & -&ig \epsilon^{ij} B_0 [E_0,[D^j, E_0]]\, , \\
    \label{eq:xizeta_app} 
  \zeta^i &=& \epsilon^{ij} \left( \left[ D^j ,E_0 [ D^l,[ D^l, B_0]]
    - B_0 [D^l,[D^l,E_0]] \right] \right.  \nonumber \\
    &-& \left. 3 [D^j,E_0][D^l,[D^l,B_0]] + 3 [D^j,B_0]
    [D^l,[D^l,E_0]] \right) \nonumber \\
    &-& 3ig E_0 [B_0, [D^i,B_0]] \, .
\end{eqnarray}

The components which we defined at order $\tau^4$ are
\begin{align}
\rho = &B_0 B_{(4)} + E_0 E_{(4)}+ \frac{1}{2}(B_{(2)} B_{(2)}  + E_{(2)} E_{(2)}) \, ,
\end{align}
\begin{align}
\kappa=&[D^i, B_0][D^i,[D^k,[D^k,B_0]]] \\
& + [D^i, E_0] [D^i,[D^k,[D^k,E_0]]  \nonumber \\
&+ ig \epsilon^{ij} [D^i, B_0] ([B_0,[D^j,B_0]]-[E_0,[D^j,E_0]]) \nonumber \\
&+ig \epsilon^{ij} [D^i, E_0] ([B_0,[D^j,E_0]]+[E_0,[D^j,B_0]]
\nonumber \\   &+[D^j,[E_0,B_0]]])   \, ,\nonumber
\end{align}
\begin{align}
\sigma = &\epsilon^{ij}[D^i, E_0][D^j,[D^k,[D^k,B_0]]] \\
&- \epsilon^{ij}[D^i, B_0][D^j,[D^k,[D^k,E_0]]] \nonumber \\
& + ig [D^i, B_0] ([B_0,[D^i,E_0]]   \nonumber \\
&+[E_0,[D^i,B_0]]+[D^i,[E_0,B_0]]) \nonumber \\
& + ig [D^i, E_0] ([E_0,[D^i,E_0]]-[B_0,[D^i,B_0]])  \, , \nonumber
\end{align}
\begin{align}
\lambda=&E^1_{(1)}E^1_{(3)} +
    B^1_{(1)}B^1_{(3)} - E^2_{(1)}E^2_{(3)} - B^2_{(1)}B^2_{(3)} \, ,
\end{align}
\begin{align}
\nu= - E^1_{(1)} E^2_{(3)} - B^1_{(1)} B^2_{(3)} -
   E^1_{(3)} E^2_{(1)} - B^1_{(3)} B^2_{(1)} \, .
\end{align}
We omit the lengthy expression for $\lambda$ and $\nu$ in terms of 
$E_0$ and $B_0$.

\section{Energy-Momentum Conservation at Order $\tau^3$}
\label{otau3}

We prove explicitly the conservation of transverse momentum at order $\tau^3$, i.e.\ the equation (\ref{eq:xi}). We do this for the first component
$\xi^1 =  \nabla^1 \left( -\triangle \varepsilon_0 + \delta - \omega \right) - \nabla^2 \gamma $, the proof for $\xi^2$ would be similar.
\begin{align}
&\nabla^1 \left( -\triangle \varepsilon_0 + \delta - \omega \right) - \nabla^2 \gamma   \nonumber \\
=&- \nabla^1 \big( (  (E_0 [ D^l,[ D^l, E_0]] + B_0 [D^l,[D^l,B_0]]) \nonumber \\
&+   ([ D^l, E_0][ D^l, E_0]+[ D^l, B_0][ D^l, B_0])  \nonumber \\
&-   ([ D^l, E_0][ D^l, E_0]+[ D^l, B_0][ D^l, B_0]) \big) \nonumber \\
&+ \frac{1}{2} \big( [D^1,[D^1,E_0]^2] + [D^1,[D^1,B_0]^2] \nonumber \\
&- [D^2,[D^2,E_0]^2] - [D^2,[D^2,B_0]^2] \big) \nonumber \\
&+ [D^2,[D^1,E_0][D^2,E_0]+[D^1,B_0][D^2,B_0]]  \nonumber \\
=& \big[D^1, E_0 [ D^l,[ D^l, E_0]] + B_0 [D^l,[D^l,B_0]]\big] \nonumber \\
&+  [ D^1, E_0][D^l, [ D^l, E_0]]+[ D^1, B_0][D^l,[ D^l, B_0]]  \nonumber \\
=&  \xi^1  \, .
\label{eq:otau3}
\end{align}
Here we have used the product rule for covariant derivatives extensively and Eq.\ (\ref{eq:id3}) for the first equal sign and Eq.\ (\ref{eq:id2}) at the second equal sign.

\end{appendix}

\end{document}